\documentclass[9pt,twocolumn]{extarticle}
\usepackage[a4paper,top=18mm,bottom=19mm,left=17mm,right=17mm,columnsep=6mm]{geometry}
\usepackage[T1]{fontenc}
\usepackage[helvratio=0.92]{newtxtext}\usepackage{newtxmath}\usepackage[scaled=0.92,varqu]{zi4}
\usepackage{microtype,booktabs,enumitem,titlesec,amsmath,tikz,xcolor,fancyhdr,caption,graphicx}
\usetikzlibrary{positioning,fit,backgrounds}
\usepackage{colortbl}
\usepackage{subcaption}
\usepackage{wrapfig}
\usepackage{placeins}
\usepackage[super,sort&compress]{natbib}
\setcitestyle{citesep={,}}
\usepackage{fontawesome5}
\usepackage[hidelinks]{hyperref}
\setlist{nosep,leftmargin=1.1em}

\definecolor{best1}{HTML}{B7E3DE}
\definecolor{best2}{HTML}{C9EAE6}
\definecolor{best3}{HTML}{DBF1EE}
\definecolor{best4}{HTML}{EDF8F6}
\definecolor{worst1}{HTML}{F2C0B8}
\definecolor{worst2}{HTML}{F5D0C9}
\definecolor{worst3}{HTML}{F8DFDA}
\definecolor{worst4}{HTML}{FBEFEC}
\usepackage{array}   % >{\raggedright\arraybackslash} column modifiers
\usepackage{multirow}

\usepackage{tabularx}
\usepackage[most]{tcolorbox}

\usepackage{listings}   % code listings for study-program figures
\definecolor{keyword}{RGB}{0,92,170}
\definecolor{comment}{RGB}{35,120,65}
\definecolor{api}{RGB}{125,55,170}
\definecolor{pykeyword}{RGB}{0,92,170}
\definecolor{pycomment}{RGB}{35,120,65}
\definecolor{pystring}{RGB}{170,55,45}
\definecolor{codeframe}{RGB}{210,210,210}
\newtcolorbox{promptbox}[1][]{
  enhanced,
  breakable,
  colback=gray!4,
  colframe=gray!45,
  boxrule=0.5pt,
  arc=1mm,
  left=1mm,right=1mm,top=1mm,bottom=1mm,
  fonttitle=\bfseries,
  #1
}

\newcommand{\titletext}{Executable Human Study Programs with Silicon Participants}

\definecolor{accent}{RGB}{204,0,0}
\DeclareCaptionLabelSeparator{bar}{\ \textbar\ }
\DeclareCaptionLabelFormat{nat}{\textbf{#1 #2}}
\titleformat{\section}{\normalsize\sffamily\bfseries}{\thesection}{0.6em}{}
\titlespacing*{\section}{0pt}{8pt}{2pt}
\titleformat{\subsection}{\small\sffamily\bfseries}{\thesubsection}{0.6em}{}
\titlespacing*{\subsection}{0pt}{6pt}{1pt}
\titleformat{\subsubsection}[runin]{\small\sffamily\bfseries}{}{0pt}{}[.]
\titlespacing*{\subsubsection}{0pt}{4pt}{0.5em}
\makeatletter\renewcommand\@biblabel[1]{#1.}\makeatother
\fancypagestyle{first}{\fancyhf{}\fancyfoot[C]{\sffamily\scriptsize\thepage}}
\renewenvironment{abstract}{\rmfamily\bfseries\small\ignorespaces}{\par}

\begin{document}
\thispagestyle{first}
\twocolumn[{\sffamily
{\color{accent}\rule{\textwidth}{1.2pt}}\par\vspace{3pt}
{\small\bfseries\color{accent}\href{https://humanstudy-hub.com/}{humanstudy-hub.com}}\par\vspace{10pt}
{\fontsize{24}{28}\selectfont\bfseries Executable Human Study Programs\\[1pt] with Silicon Participants\par}
\vspace{8pt}
{\small Xuan Liu\textsuperscript{1,2,\,\scriptsize\faEnvelope[regular]}, Haoyang Shang\textsuperscript{1,3}, Yuanjun Feng\textsuperscript{1,4}, Shao-Yu Chu\textsuperscript{2},
Qiyu Li\textsuperscript{2}, Yash Raj Shrestha\textsuperscript{4}, Haojian Jin\textsuperscript{2}}\par\vspace{3pt}
{\scriptsize \textsuperscript{1}\href{https://humanstudy-hub.com/}{HumanStudy Hub}\quad \textsuperscript{2}University of California San Diego\quad
\textsuperscript{3}University of British Columbia\quad \textsuperscript{4}University of Lausanne}\par\vspace{2pt}
{\scriptsize \faEnvelope[regular]\ \href{mailto:xul049@ucsd.edu}{xul049@ucsd.edu}}\par\vspace{10pt}
\begin{minipage}{\textwidth}\begin{abstract}

We introduce \textsc{HumanStudyProgram}, a framework that represents a documented human study as an executable program with configurable LLM-based silicon participants. Its computational abstraction separates study logic from participant implementation through standardized APIs for presenting information and collecting responses. We implement a researcher-facing authoring tool that reconstructs source-linked study descriptions from papers and supplementary materials, supports researcher review, and generates executable \textsc{HumanStudyProgram} artifacts. Researchers can then configure silicon participants, run the programs, and analyze results from aggregate findings to individual responses. We demonstrate the approach through study cases and evaluate the tool through a hands-on expert review with 12 social science researchers. All completed the workflow and identified potential uses in understanding published protocols, prototyping studies, and examining AI behavior. We aim to improve the reproducibility of human studies by reconstructing their documented procedures as programs that others can reuse and build upon.

\end{abstract}
\end{minipage}\vspace{10pt}
\begin{minipage}{\textwidth}\rmfamily\centering
\includegraphics[width=\textwidth]{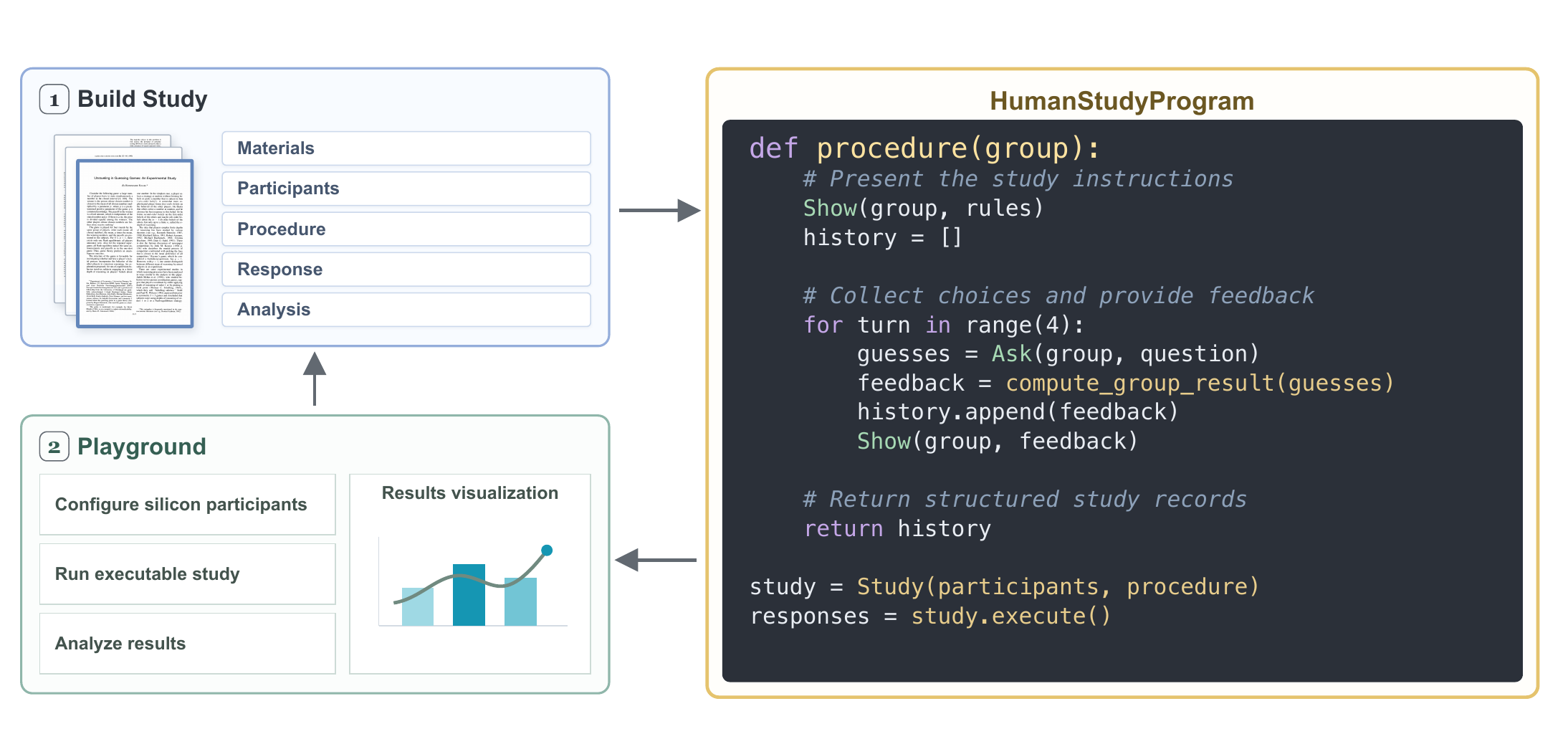}
    \captionof{figure}{\textbf{The \textsc{HumanStudyProgram} Artifact and Its Authoring Tool.} A \textsc{HumanStudyProgram} represents a human study as an executable program: \texttt{Show} and \texttt{Ask} express participant-facing interactions, while ordinary Python expresses the study logic connecting them. We implement this abstraction in an authoring tool with two linked stages. \textcircled{1} \textit{Build Study} generates a \textsc{HumanStudyProgram} from a paper and its supplementary materials. \textcircled{2} \textit{Playground} configures silicon participants, runs the generated \textsc{HumanStudyProgram}, and analyzes its results. The two stages form an iterative workflow for generating and running \textsc{HumanStudyProgram}s.}
    
    \label{fig:overview}
\end{minipage}\vspace{12pt}}]

\section{Introduction}
\label{sec:intro}

Major HCI venues (e.g., CHI) now encourage authors to provide study materials and ask reviewers to consider their availability~\citep{acmArtifactReview,chi2026submissionguide}. The proportion of CHI papers that shared interview protocols increased from 2\% in 2017 to 25\% in 2022~\citep{chakravorti2025embracing,niksirat2023changes}.
However, using these materials still requires substantial effort~\citep{iarygina2026computational}.
First, researchers must recover the exact wording, task order, and condition assignments from information scattered across the methods section, figures, appendices, and supplementary materials~\citep{niksirat2023changes,vanderdonckt2025context}.
Second, researchers have to reconstruct the complete protocol from high-level descriptions, such as ``tasks were counterbalanced in order and randomized across conditions''. Third, to test these protocols, researchers often need to secure ethics approval, recruit and compensate participants, and administer the study.  

In contrast, empirical machine learning (EML) has benefited from an infrastructure for frictionless reproducibility~\citep{donoho2023data}, in which researchers can build directly on one another’s digital artifacts. Researchers routinely release repositories containing datasets, model weights, training code, and evaluation scripts~\citep{wolf-etal-2020-transformers,lhoest-etal-2021-datasets}. To build on a published result, a researcher can clone its repository, reproduce the reported findings, and inspect the code to diagnose discrepancies. Once the baseline is established, they can modify individual components—substituting a dataset, adjusting a hyperparameter, or replacing a model module—and rerun the pipeline to observe the effects. Each iteration requires only a fraction of the effort needed to reconstruct the original work, while shared repositories accumulate fixes and extensions contributed by others. This capacity for rapid verification and iteration is widely regarded as a key driver of AI’s striking progress~\citep{donoho2023data}.

This paper presents \textsc{HumanStudyProgram}, a researcher-facing framework that reconstructs human study protocols documented in papers and supplementary materials as executable programs and interfaces the programs with configurable LLM-based silicon participants through standardized APIs (Figure~\ref{fig:overview}). The resulting program is a reusable artifact that researchers can inspect and modify. Running it with silicon participants generates structured synthetic response data without recruiting a new human sample, allowing researchers to exercise the procedure and analysis end to end, identify implementation problems, and prototype alternative questions or conditions. \textsc{HumanStudyProgram} targets interviews, surveys, and in-lab experiments whose essential procedures can be represented as exchanges of information and responses.
These agent-based runs complement human-participant research by supporting protocol inspection and early prototyping~\citep{anthis2025position}; they do not replace human evidence for claims about people~\citep{participants_CHI,participants_qualitative_CHI,wang2026interviewinformed,gui2026peeking}.

At the core of \textsc{HumanStudyProgram} is our observation that diverse human study protocols
share a common interaction structure: they present information, collect responses, and use
procedural logic to determine what happens next. This observation allows us to \textbf{separate study
logic from the mechanisms used to interact with participants}. We introduce two API operations 
to realize this separation of concerns for participants: \texttt{Show} presents information to one or more participants,
while \texttt{Ask} poses a question and returns their responses (Figure~\ref{fig:study-abstraction}(b)). A \textsc{HumanStudyProgram} uses these
operations for all participant-facing interactions and ordinary Python for condition
assignment, control flow, state updates, computation, and coordination. The protocol therefore specifies what participants receive and return without depending on how a participant is implemented. This separation supports studies ranging from single-question surveys to multi-round, multi-participant experiments and allows researchers to vary silicon participants
while holding the protocol fixed.

Behind \texttt{Show} and \texttt{Ask}, the runtime instantiates each configured
agent, assembles its context, invokes the model, parses its response, and manages its memory
across interactions. 
Agent memory is enabled by default. \texttt{Show} adds presented information to the participant's
context so that later interactions can refer to it. By default, \texttt{Ask} similarly records
both its question and the returned response after invoking the configured model. A study can
instead call \texttt{Ask} with \texttt{remember=False} to collect a response without adding that
exchange to subsequent context. Such probes still read existing memory; only the new
question and response are excluded. These
semantics support both isolated measurements and stateful, multi-round interactions without
requiring each study program to implement agent memory or model-specific prompting.

We implement an authoring tool for creating and running \textsc{HumanStudyProgram}
artifacts through two linked stages (Section~\ref{sec:system-reorganized}). In \emph{Build Study}, an
LLM-assisted pipeline extracts Materials, Participants, Procedure, Response, and Analysis from
a paper and its supplementary files, links each component to its source, and flags missing,
unclear, or conflicting information for researcher review. Once the researcher approves this description, the system generates
an executable package containing the confirmed study materials, participant definitions, a
Python procedure module, and a measurement module for the analysis; it then validates
the code and performs a simulated trial run. In \emph{Playground}, researchers configure the
model, participant specification, memory, population, and condition assignments, execute the
generated package, and inspect results from aggregate findings to individual interactions. The
runtime records each agent's configuration, visible context, raw output, parsed response, and
resulting study state.

We validate the implementation through study demonstrations and evaluate its use through
a hands-on expert review. The demonstrations use four cases from three cognitive, strategic,
and social-psychology studies to examine recovery of missing information, execution of a
multi-round group protocol, transparency into agent context, and navigation from findings to
individual responses. In the expert review, 12 doctoral-level social science researchers each
selected a familiar study, corrected and approved its reconstructed description, configured and ran silicon participants, and
interpreted the results. All 12 completed the workflow, collectively reviewing 98 flagged
issues across their study packages. They identified value in understanding published
protocols, prototyping studies before human data collection, and applying established
protocols to study AI behavior, while emphasizing that simulated results cannot substitute for
human evidence.

This paper makes three contributions. First, it introduces a computational abstraction for
human studies that combines participant-facing \texttt{Show} and \texttt{Ask} operations with
ordinary program logic, separating an executable protocol from the agents used to run it.
Second, it contributes an authoring tool for creating and executing
\textsc{HumanStudyProgram} artifacts, supporting source-linked protocol reconstruction,
researcher review and approval, configurable silicon participants, and analysis from aggregate
findings to individual responses. Third, it provides evidence from study demonstrations
and a hands-on review with 12 social science researchers, showing the abstraction's ability to
represent varied study procedures and identifying its value for understanding
protocols, prototyping studies, and examining AI behavior.

\section{Background and Related Work}
We review prior work on reproducibility and reuse in human-subject research, tools for authoring and running human studies, and LLMs in behavioral research.

\subsection{Reproducibility and Reuse in Human-Subject Research}
Efforts to improve the reproducibility of human-subject research initially centered on conducting more direct replications (Table~\ref{tab:reuse-progression}). In HCI, RepliCHI called for the community to replicate and validate prior findings more systematically~\citep{wilson2011replichi}. Large collaborative projects subsequently coordinated replications across laboratories and participant samples, including Many Labs, the Reproducibility Project: Psychology, and replications of social science experiments published in \textit{Nature} and \textit{Science}~\citep{klein2014investigating,open2015estimating,camerer2018evaluating}. These projects made replicability empirically assessable, but each study still had to be recovered from the published record, reimplemented, and run with newly recruited participants. They therefore demonstrated both the value of reuse and the substantial coordination and execution costs it entails.

\begin{table*}[t]
    \centering
    \footnotesize
    \setlength{\tabcolsep}{4pt}
    \renewcommand{\arraystretch}{1.15}
    \caption{Historical progression toward reusable human-subject research.}
    \label{tab:reuse-progression}
    \begin{tabularx}{\textwidth}{@{}>{\raggedright\arraybackslash}p{0.11\textwidth}>{\raggedright\arraybackslash}p{0.20\textwidth}>{\raggedright\arraybackslash}p{0.25\textwidth}>{\raggedright\arraybackslash}X@{}}
        \toprule
        \textbf{Period} & \textbf{Approach} & \textbf{Reusable object} & \textbf{Examples} \\
        \midrule
        2011--2018 & Direct replication & Study findings with newly recruited participants & RepliCHI~\citep{wilson2011replichi}; Many Labs~\citep{klein2014investigating}; Reproducibility Project: Psychology~\citep{open2015estimating}; Camerer et al.~\citep{camerer2018evaluating} \\
        \arrayrulecolor{black!20}\cmidrule(lr){1-4}\arrayrulecolor{black}
        2017--2023 & Artifact sharing & Study plans, materials, data, and analysis code & Open Science Framework~\citep{foster2017open}; HARK No More~\citep{cockburn2018hark}; CHI Artifact Transparency~\citep{wacharamanotham2020transparency}; ACM Artifact Badging~\citep{acmArtifactReview}; Living Papers~\citep{heer2023living} \\
        \arrayrulecolor{black!20}\cmidrule(lr){1-4}\arrayrulecolor{black}
        2025--2026 & Agentic reproduction & Existing computational experiments and analyses & CORE-Bench~\citep{siegel2024corebench}; Paper2Code~\citep{seo2025paper2code}; ReplicatorBench~\citep{nguyen2026replicatorbench} \\
        \bottomrule
    \end{tabularx}
\end{table*}

The next wave of work sought to make future studies easier to verify and reuse by preserving their plans and artifacts. The Open Science Framework provided infrastructure for registering study plans and sharing protocols, stimuli, data, and code~\citep{foster2017open}; preregistration was subsequently advocated within CHI to commit hypotheses and analyses before results were known~\citep{cockburn2018hark}. Artifact surveys and ACM's review and badging framework then shifted attention from whether materials were shared to whether they were documented, functional, and reusable~\citep{wacharamanotham2020transparency,iarygina2026computational,acmArtifactReview}. More recently, Living Papers demonstrated how scholarly documents can incorporate executable code and interactive components rather than remain static accounts~\citep{heer2023living}. Together, these efforts make research outputs increasingly available, structured, and interactive. However, they support the reuse of artifacts and documents rather than making the participant-facing study itself rerunnable.

Recent work has begun to use agents to support the reproduction of published experiments~\citep{kim2026autoexperiment,hu2025reprobench,pagel2026verification}. CORE-Bench tests whether agents can re-execute computational experiments using paper-associated code and data~\citep{siegel2024corebench}; Paper2Code generates implementations from machine-learning papers~\citep{seo2025paper2code}; and ReplicatorBench extends automated reproduction to statistical analyses reported in social science papers~\citep{nguyen2026replicatorbench}. These computational approaches either target computational studies or, when applied to human-subject research, reproduce only the post-data-collection analysis over existing data. \textsc{HumanStudyProgram} extends reuse to the study procedure itself, reconstructing human studies as programs that can generate new responses through silicon participants.

\subsection{Tools for Authoring and Running Human Studies}

Behavioral experiment software has tried to make human studies programmatically executable. Tools such as PsychoPy, jsPsych, and lab.js support the implementation of experimental procedures as sequences of stimuli, trials, and response collection~\citep{peirce2019psychopy2,deleeuw2015jspsych,henninger2022labjs}, while experiment repositories have packaged such implementations for reuse~\citep{sochat2016experiment}. These tools provide precise control over experimental delivery, but their abstractions primarily center on individual human interactions: presenting stimuli through an interface and capturing responses through input devices.

Later platforms extended executable experimentation to multi-participant settings. z-Tree, oTree, Empirica, and LIONESS Lab support rounds, roles, information exchange, and payoff structures required by social and economic experiments~\citep{fischbacher2007ztree,chen2016otree,almaatouq2021empirica,giamattei2020lioness}. These platforms help researchers implement studies they plan to conduct. \textsc{HumanStudyProgram} instead turns studies already described in papers and supplementary materials into executable programs, without requiring researchers to implement each study from scratch.

\subsection{LLMs in Behavioral Research}

Researchers use LLMs as silicon participants both to model human behavior and to study AI behavior in its own right. In the former role, agents are used to reproduce findings from human studies, model the responses of particular populations, and benchmark agent--human correspondence across tasks and populations~\citep{ashokkumar2026predict,hamalainen2023synthetic,argyle2023out,horton2023large_eco,aher2023using,hu2025simbench,xie2026ssdatabench,Bisbee_Clinton_Dorff_Kenkel_Larson_2024}. In the latter, established human study paradigms provide controlled settings for examining how AI agents perceive, decide, and interact~\citep{rahwan2019machine,binz2023using,yax2024reasoning,strachan2024theory,akata2025games}. Whether agent--human resemblance is necessary therefore depends on the claim being made. Claims about people require correspondence to be empirically established rather than assumed~\citep{hullman2026human,participants_CHI,participants_qualitative_CHI,participants_nature,wang2026interviewinformed,gui2026peeking}; when AI behavior itself is the object of inquiry, both correspondence with and departure from human findings are informative. Across these purposes, silicon participants have been used in individual surveys and behavioral tasks, as well as multi-agent simulations of social interaction, collective behavior, and group coordination~\citep{ashkinaze2025plurals,park2023generativeagentsinteractivesimulacra,vezhnevets2023generativeagentbasedmodelingactions,piao2025agentsocietylargescalesimulationllmdriven}.

Another line of work explores how silicon participants should be specified, from demographic personas and autobiographical backstories~\citep{park2024generativeagentsimulations1000,lutz-etal-2025-prompt,wang2026interviewinformed} to behavioral fine-tuning~\citep{binz2025foundation,kolluri2025finetuningllmshumanbehavior}. Empirical studies show that simulated behavior depends not only on the underlying model, but also on persona conditioning, prompt phrasing, option ordering, and response format~\citep{hu2024quantifyingpersonaeffectllm,pmlr-v202-santurkar23a,tjuatja2024llmsexhibithumanlikeresponse,dominguez-olmedo2024questioning}. A simulated response must therefore be interpreted together with the participant configuration and experimental context that produced it~\citep{pilli2026emulating}.

\textsc{HumanStudyProgram} makes human study protocols available as reusable programs for investigating AI behavior, supporting the research directions reviewed above. Researchers can vary the study procedure and participant configuration independently and trace results to the interactions that produced them.

\newcommand{\introrowsep}{%
  \arrayrulecolor{black!20}%
  \cmidrule(lr){1-2}%
  \arrayrulecolor{black}%
}

\begin{table*}[t]
\centering
\footnotesize
\setlength{\tabcolsep}{4pt}
\renewcommand{\arraystretch}{1.08}
\caption{Operations supported by \textsc{HumanStudyProgram} study API and by Python.}
\label{tab:intro-study-operations}

\begin{tabularx}{\textwidth}{
    @{}
    >{\raggedright\arraybackslash}p{0.35\textwidth}
    >{\raggedright\arraybackslash}X
    @{}
}
\toprule
\textbf{Supported operation} & \textbf{Example} \\
\midrule

\multicolumn{2}{@{}l}{\textbf{Provided by the study API}} \\
\introrowsep

Define a study.
&
\texttt{Study(participants, procedure) -> Study}
\\
\introrowsep

Execute a study with silicon participants.
&
\texttt{study.execute() -> T};
raises \texttt{StudyExecutionError} on failure.
\\
\introrowsep

Present content to silicon participants.
&
\texttt{Show(participants, content) -> None};
raises \texttt{ParticipantInteractionError} on failure.
\\
\introrowsep

Collect responses from silicon participants.
Pass \texttt{options} for a closed response set, and
\texttt{remember=False} to avoid updating memory.
&
\texttt{Ask(participants, question, options=None,}
\newline
\texttt{remember=True)}
\newline
\texttt{-> dict[ParticipantID, Response]};
raises \texttt{ParticipantInteractionError} on failure.
\\

\midrule
\multicolumn{2}{@{}l}{\textbf{Provided by Python}} \\
\introrowsep

Select participants.
&
\texttt{group = [p for p in participants if p.role == "investor"]}
\\
\introrowsep

Apply configured roles or conditions.
&
\texttt{p.condition = assignments[p.id]}
\\
\introrowsep

Sequence activities.
&
\texttt{Show(group, tutorial); answers = Ask(group, question)}
\\
\introrowsep

Repeat activities or rounds.
&
\texttt{for round in range(4): ...}
\\
\introrowsep

Branch the procedure.
&
\texttt{if answer == "yes": ...}
\\
\introrowsep

Maintain and update study state.
&
\texttt{state = \{"round": 1\}; state["round"] += 1}
\\
\introrowsep

Perform calculations.
&
\texttt{target = (2 / 3) * mean(guesses.values())}
\\
\introrowsep

Represent study materials.
&
\texttt{question = "Choose a number from 0 to 100"}
\\
\introrowsep

Pass information between participants.
&
\texttt{Show(trustees, transfers)}
\\
\introrowsep

Construct returned study data.
&
\texttt{return \{"guesses": guesses, "target": target\}}
\\
\introrowsep

Handle failures.
&
\texttt{try: responses = study.execute()}
\\
\introrowsep

Run activities concurrently.
&
\texttt{responses = await asyncio.gather(*tasks)}
\\
\introrowsep

Analyze returned data.
&
\texttt{results = analyze(responses)}
\\

\bottomrule
\end{tabularx}
\end{table*}

\section{Design Process}
\label{sec:design-goals}
The design of \textsc{HumanStudyProgram} drew on three complementary
activities: examination of published study procedures, informal discussions
with social science researchers, and iterative implementation of study
programs. These activities addressed two design questions: which operations
are needed to express different study procedures, and how to support
researchers in reconstructing, executing, and interpreting those procedures
with silicon participants.

\noindent\textbf{Understanding Study Procedures.} 
We reviewed more than 50 papers reporting human-subject studies. We focused on
social experiments designed to test theories through participant behavior.
For each study, we read the paper and available supplementary materials and
traced what participants were shown, what responses were requested, how the
study state changed, and how the responses were used in subsequent
interactions or analysis. We also considered how conditions and participant
roles altered this sequence. These relationships informed the five-part study
description: Materials, Participants, Procedure, Response, and Analysis.

Across these studies, we found a stable distinction between interactions with
participants and the program logic that organizes those interactions. We began
with an initial API and refined it while implementing 13 studies spanning
single-response tasks, multistage procedures, repeated interactions with
feedback, role-based multi-participant experiments, and studies containing
multiple empirical substudies. Through this process, we developed two
participant-facing operations: \texttt{Show}, which presents information, and
\texttt{Ask}, which poses a question and collects responses. Condition
assignment, branching, computation, state updates, and participant coordination
remain expressible through ordinary program logic. These abstractions are
formalized in Section~\ref{sec:study-abstractions}.

\noindent\textbf{Understanding Researcher Needs.}
Over more than six months, we held informal discussions with approximately
30 doctoral-level social science researchers who conduct human studies,
primarily in psychology, management, economics, and sociology. We discussed
how they reconstruct published procedures, configure study runs, and interpret
results, and used their feedback while iterating on the tool. Through this
process, we arrived at three considerations for the surrounding workflow:
source-grounded reconstruction, configurable and inspectable execution, and
traceable interpretation.

For reconstruction, these discussions motivated us to support comparison of
extracted study information with its sources and resolution of missing,
inconsistent, or incorrectly extracted details before execution. We therefore
designed a review interface that places study information alongside source
evidence and requires approval of the reconstructed description.

For execution, researchers needed to configure silicon participants
individually or as groups, assign them to conditions or roles, and see the
complete context provided to each participant. They also needed to control
participant memory: context should carry across interactions when required by
the study, while isolated measurements should not affect later responses. We
therefore separated study programs from reusable participant and group
configurations, made execution context visible, and supported interaction-level
memory control through \texttt{remember=False}.

For interpretation, the discussions motivated support for tracing aggregate
findings back to condition-level results, analysis outputs, and individual
responses. We designed linked views across these levels to help researchers
examine how a reported result was produced. Together, these design choices
connect source review, participant configuration, and result inspection to
the executable study abstraction; Section~\ref{sec:system-reorganized}
describes the corresponding interfaces.

\begin{figure*}[t]
\centering
\includegraphics[width=\textwidth]{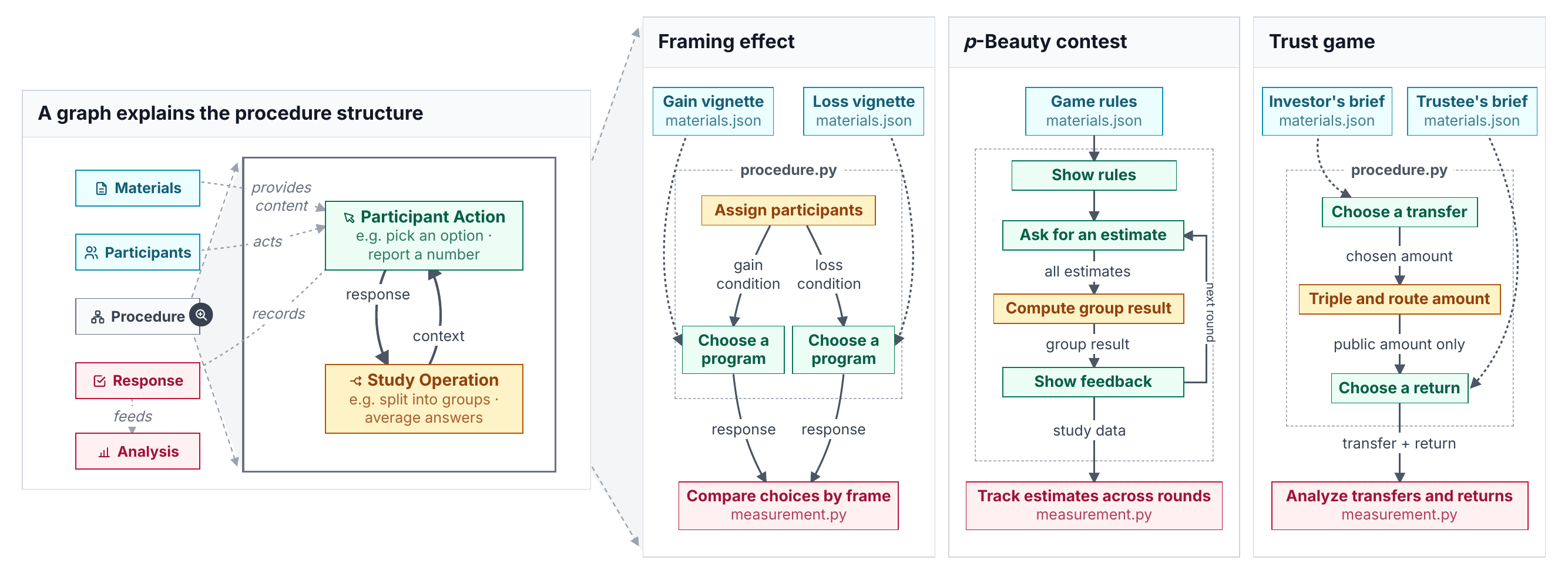}
\caption{\textbf{Representing Study Procedure Structure.} The five-part
study description situates Procedure within Materials, Participants, Response,
and Analysis (left). A graph illustrates Procedure using \emph{Participant
Actions} and \emph{Study Operations}, with edges carrying responses and
context between them. Different compositions express independent conditions,
repeated group feedback, and information exchange between participant roles
(right).}
\label{fig:computational-model}
\end{figure*}

\section{\textsc{HumanStudyProgram} Abstractions}
\label{sec:study-abstractions}

Our design process identified participant-facing interactions and study logic
as a recurring structure across study procedures. We formalize that structure
here and show how it becomes a program.

\subsection{Study and Participant Abstractions}
\label{subsec:study-abstraction}

A \textsc{HumanStudyProgram} represents a study through five connected parts:
Materials, Participants, Procedure, Response, and Analysis. Materials contain
the information presented in the original study, and Participants describe its
reported population and roles. Procedure determines how the study unfolds and
produces data structured by Response; Analysis defines how those data are
transformed into study findings. Figure~\ref{fig:computational-model} situates
Procedure within these five parts.

Within Procedure, we distinguish between interactions with participants and
the study logic that organizes those interactions. A \emph{Participant Action}
invokes one or more participants to present information or collect a response.
A \emph{Study Operation} acts on the study without invoking a participant, for
example by assigning conditions, performing computations, updating study
state, or routing information. For explanatory purposes, we illustrate a
procedure as a graph in which edges carry responses from Participant Actions
to Study Operations and return the resulting context to later Participant
Actions. The executable procedure itself remains an ordinary Python program.
Figure~\ref{fig:computational-model} shows how these operations compose to
express independent conditions in the Framing Effect~\citep{Tversky1981TheFO},
repeated group feedback in the $p$-Beauty Contest~\citep{Selten2007UnravelingIG},
and information exchange between roles in the Trust Game~\citep{Berg1995TrustRA}.

Participant Actions operate on silicon participants that persist throughout a
study run. Each participant maintains a model configuration, a participant
specification, and private memory that is enabled by default across
interactions. Figure~\ref{fig:intro-participant-abstraction}(a) shows this
participant representation and how an interaction may update or leave its
memory unchanged.

We implement Participant Actions through two study API operations:
\texttt{Show} presents information to one or more participants, and
\texttt{Ask} poses a question and returns their responses. Ordinary Python
expresses the surrounding study logic. The \texttt{Study} abstraction binds a
configured set of participants to a procedure and executes that procedure.
Figure~\ref{fig:study-abstraction}(b) shows these relationships in code, and
Table~\ref{tab:intro-study-operations} summarizes the operations provided by
the study API and by Python.

\lstdefinestyle{introabstraction}{
  language=Python,
  basicstyle=\ttfamily\scriptsize,
  keywordstyle=\color{keyword}\bfseries,
  commentstyle=\color{comment}\itshape,
  morekeywords=[2]{Participant,Study,Ask,Show,generate},
  keywordstyle=[2]\color{api}\bfseries,
  frame=single,
  framerule=0.3pt,
  breaklines=true,
  columns=fullflexible,
  keepspaces=true,
  showstringspaces=false,
  aboveskip=1pt,
  belowskip=0pt,
  lineskip=-0.5pt
}

\begin{figure*}[t]
\centering
\begin{minipage}[t]{0.485\textwidth}
\textbf{(a) Participant}
\begin{lstlisting}[style=introabstraction]
class Participant:
    # Configure one silicon participant for a run.
    def __init__(self, id, model, specification,
                 memory_enabled=True):
        self.id = id  # Keys this participant's responses.
        self.model = model  # Model and inference settings.
        self.specification = specification  # Participant traits.
        # Memory is private and enabled by default.
        self.memory = [] if memory_enabled else None

    def show(self, content):
        # Retain presented content when memory is enabled.
        if self.memory is not None:
            self.memory.append({"type": "observation",
                                "content": content})

    def inquiry(self, question, options=None):
        # Generate a response and retain the exchange.
        response = generate(self.model, self.specification,
                            self.memory or [], question, options)
        if self.memory is not None:
            self.memory.append({"type": "interaction",
                                "question": question,
                                "response": response})
        return response

    def interrogate(self, question, options=None):
        # Generate a response without changing memory.
        return generate(self.model, self.specification,
                        self.memory or [], question, options)
\end{lstlisting}
\end{minipage}
\hfill
\begin{minipage}[t]{0.485\textwidth}
\textbf{(b) Study}
\begin{lstlisting}[style=introabstraction]
def Show(participants, content):
    # Present the same content to a group.
    for p in participants:
        p.show(content)  # Retained when memory is enabled.

def Ask(participants, question, options=None,
        remember=True):
    # Collect one response per participant, keyed by id.
    responses = {}
    for p in participants:
        if remember:
            # Retain the question and response.
            responses[p.id] = p.inquiry(question, options)
        else:
            # Collect a probe without changing memory.
            responses[p.id] = p.interrogate(question, options)
    return responses

class Study:
    # Bind a configured population to a procedure.
    def __init__(self, participants, procedure):
        self.participants = participants
        self.procedure = procedure

    def execute(self):
        # Run the procedure and return its output.
        return self.procedure(self.participants)

study = Study(participants, procedure)
results = study.execute()
\end{lstlisting}
\end{minipage}
\vspace{6pt}
\caption{\textbf{Participant and Study Abstractions.} Example code for the two
abstractions. In (a), a \texttt{Participant} combines a model and specification with memory enabled by default;
\texttt{inquiry} retains an interaction when memory is enabled, whereas
\texttt{interrogate} does not. In (b), \texttt{Show} and
\texttt{Ask} apply these operations to a group, and \texttt{Study} binds that
group to an executable procedure. In Figure~\ref{fig:overview}, the highlighted \texttt{procedure} calls these \texttt{Show} and \texttt{Ask} operations; \texttt{Study.execute} invokes the function with the configured participants and returns its records. The code omits runtime details.}
\label{fig:study-abstraction}
\label{fig:intro-participant-abstraction}
\end{figure*}

\begin{figure*}[t]
\centering
\includegraphics[width=\textwidth]{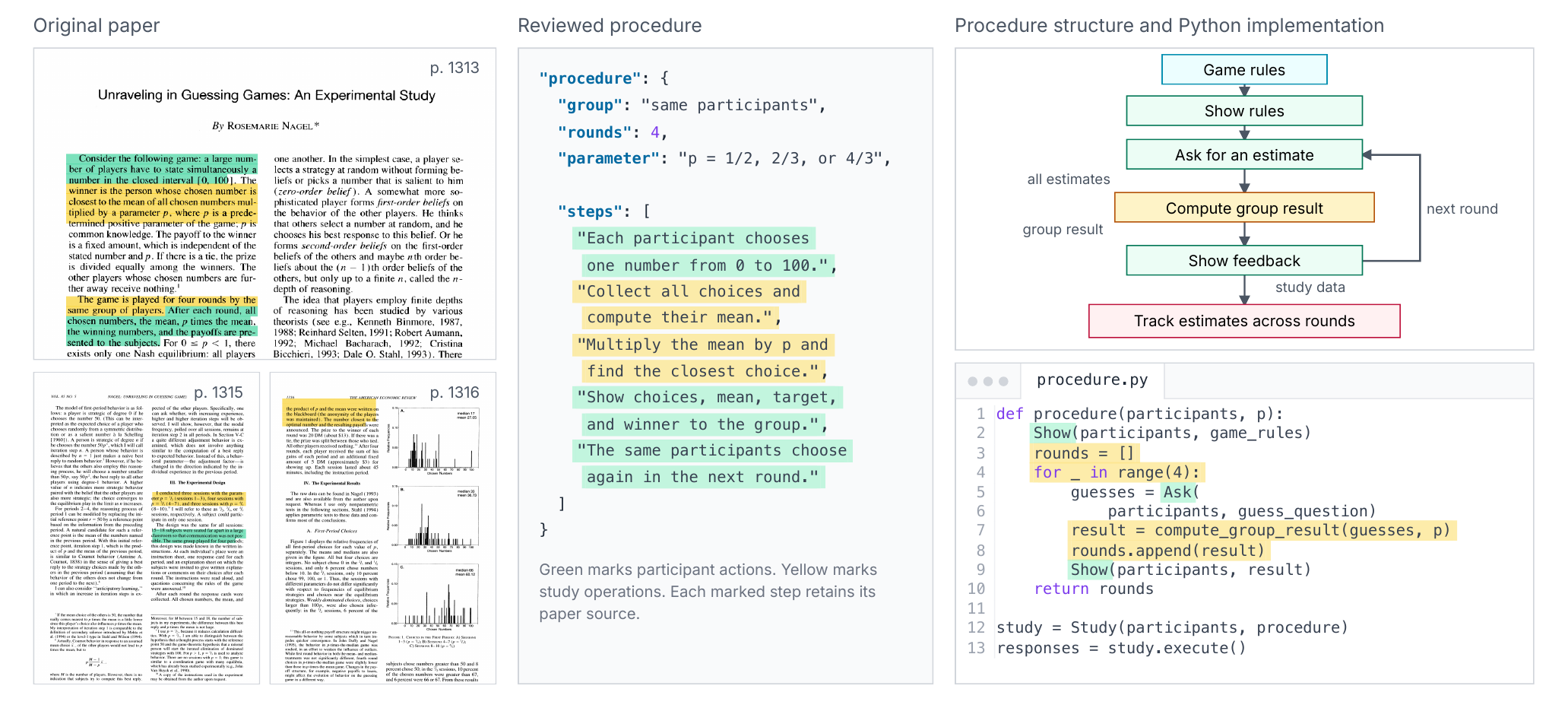}
\caption{\textbf{Representing a Multi-Round Study Procedure.} The four-round
$p$-Beauty Contest is transformed from source passages into a reviewed
procedure, an explanatory graph, and an executable Python program.}
\label{fig:case-beauty}
\end{figure*}

\subsection{Expressing Different Study Procedures}
\label{subsec:example-programs}

\lstdefinestyle{study}{
  language=Python,
  basicstyle=\ttfamily\scriptsize,
  keywordstyle=\color{pykeyword}\bfseries,
  commentstyle=\color{pycomment}\itshape,
  stringstyle=\color{pystring},
  morekeywords=[2]{Study,Ask,Show},
  keywordstyle=[2]\color{api}\bfseries,
  frame=single,
  rulecolor=\color{codeframe},
  framerule=0.6pt,
  breaklines=true,
  columns=fullflexible,
  keepspaces=true,
  showstringspaces=false,
  aboveskip=2pt,
  belowskip=2pt,
  xleftmargin=2pt,
  xrightmargin=2pt
}

\newcommand{\programtitle}[1]{%
  \noindent\textbf{#1}\vspace{-2pt}
}

\begin{figure*}[t]
\centering

\begin{minipage}[t]{0.325\textwidth}
\programtitle{(a) Between-subject framing study}
\begin{lstlisting}[style=study,basicstyle=\ttfamily\tiny]
def procedure(participants):
    gain, loss = assign_groups(
        participants, assignment_config)
    groups = {
        "gain": gain, "loss": loss}
    questions = {
        "gain": gain_question,
        "loss": loss_question}
    return {
        condition: Ask(
            group, questions[condition])
        for condition, group
        in groups.items()}
\end{lstlisting}
\end{minipage}
\hfill
\begin{minipage}[t]{0.325\textwidth}
\programtitle{(b) Anchoring study}
\begin{lstlisting}[style=study,basicstyle=\ttfamily\tiny]
def procedure(participants):
    low, high = assign_groups(
        participants, assignment_config)
    responses = {}
    for item in items:
        conditions = [
            (low, item.low),
            (high, item.high)]
        for group, anchor in conditions:
            comparison = Ask(
                group, item.compare(anchor))
            estimate = Ask(
                group, item.estimate)
            responses[(item.id, anchor)] = {
                "comparison": comparison,
                "estimate": estimate}
    return responses
\end{lstlisting}
\end{minipage}
\hfill
\begin{minipage}[t]{0.325\textwidth}
\programtitle{(c) Trust Game}
\begin{lstlisting}[style=study,basicstyle=\ttfamily\tiny]
def procedure(participants):
    investors = [
        p for p in participants
        if p.role == "investor"]
    trustees = [
        p for p in participants
        if p.role == "trustee"]
    outcomes = []
    pairs = zip(investors, trustees)
    for investor, trustee in pairs:
        sent = Ask(
            [investor],
            transfer_question)[investor.id]
        available = 3 * sent
        returned = Ask(
            [trustee],
            return_question(sent, available)
        )[trustee.id]
        outcomes.append({
            "investor_id": investor.id,
            "trustee_id": trustee.id,
            "sent": sent,
            "returned": returned})
    return outcomes
\end{lstlisting}
\end{minipage}

\vspace{-2pt}
\caption{\textbf{Additional Example Study Programs.} Three further procedures:
a between-subject framing study, an anchoring study, and the Trust Game.
Participant interactions use
\texttt{Ask}; condition assignment, iteration, computation, and information
routing use Python. The illustrative \texttt{assign\_groups} helper applies the
assignment configured in Playground, including random allocation when selected;
it is not an additional study API operation.}
\label{fig:example-study-programs}
\end{figure*}

We first use a stateful, multi-round procedure to show how Participant Actions
and Study Operations work together in an executable program.
In the $p$-Beauty Contest~\citep{Selten2007UnravelingIG}, each participant chooses a number between 0 and 100,
and the participant whose choice is closest to $p$ times the group mean wins.
The study repeats this process for four rounds and shows the group result after
each round, so later choices are made with the results of earlier rounds.
Figure~\ref{fig:case-beauty} traces these repeated choices, group calculation,
and feedback from their source passages to a reviewed procedure and its Python
implementation. Participant Actions collect estimates through \texttt{Ask}
and present group feedback through \texttt{Show}; Study Operations compute
the group result and control the repeated rounds.

More examples show how the same operations express procedures with different
structures. Figure~\ref{fig:example-study-programs} uses independent condition
assignment in a framing study~\citep{Tversky1981TheFO}, repeated item and
condition assignment in an anchoring study~\citep{Jacowitz1995MeasuresOA}, and
information transfer between roles in the Trust Game~\citep{Berg1995TrustRA}.

\section{\textsc{HumanStudyProgram} Runtime}
\label{sec:runtime}

The abstractions in Section~\ref{sec:study-abstractions} define a study
program and the silicon participants that execute it. We now describe the
supporting infrastructure: how a study is packaged for execution, how the
runtime carries out its procedure, and how study data and execution traces are
recorded for analysis.

\subsection{Generating an Executable Study Package}
\label{subsec:generate-package}

\begin{figure*}[t]
\centering
\includegraphics[width=\textwidth]{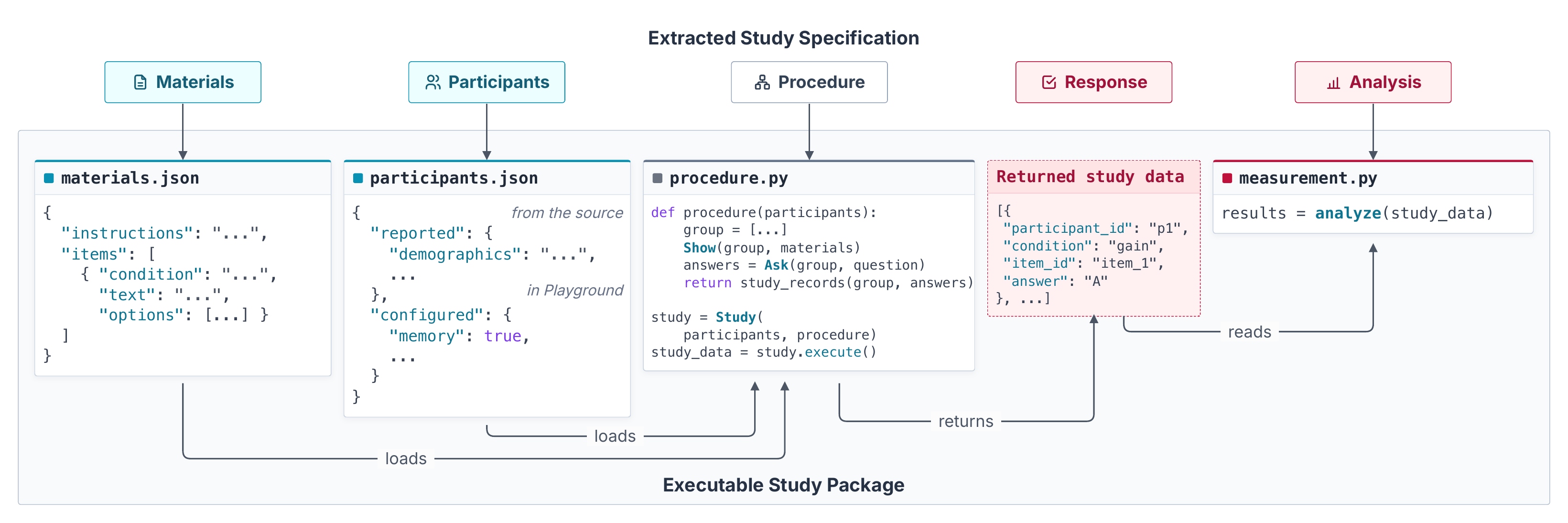}
\caption{\textbf{Executable Study Package.} A confirmed five-part study
description becomes the files loaded by the shared runtime. Materials and
Participants configure Procedure in \texttt{procedure.py}, Response defines the
study data returned by Procedure, and \texttt{measurement.py} implements
Analysis.}
\label{fig:study-package-reorganized}
\end{figure*}

Once researchers approve the five-part study description, the tool
generates the executable study package shown in
Figure~\ref{fig:study-package-reorganized}.
Confirmed Materials are stored in \texttt{materials.json}.
\texttt{participants.json} retains the reported population, roles, and
attributes, together with fields for the silicon-participant configuration
used in a run. From the confirmed Procedure, the tool generates
\texttt{procedure.py}, which uses the study API for participant interactions and
ordinary Python for study-specific logic. The procedure module organizes participant
responses and computed outcomes according to the Response specification, and
\texttt{measurement.py} implements Analysis over the resulting study data.
The worked example in Section~\ref{subsec:example-programs} illustrates this
program structure.

Before making the package available for execution, the tool analyzes the
generated code structure and conducts a safe test run to identify invalid
operations and failures in the study logic. Packages that pass these checks
are made available to the runtime.

\subsection{Executing Study Programs and Recording Results}
\label{subsec:runtime-execution}

At runtime, the system instantiates the configured participants and executes
Procedure in \texttt{procedure.py}. Memory is enabled by default, so
content presented by \texttt{Show} and interactions collected by
\texttt{Ask} become available to later participant calls. To interview or
probe a participant without changing its subsequent context, a study calls
\texttt{Ask(..., remember=False)}. This operation reads existing memory but
does not retain the new question or response.
Independent calls can run in parallel, whereas interactive and multi-round
procedures wait for the responses needed to continue. For example, each round
of the $p$-Beauty Contest in Figure~\ref{fig:case-beauty} collects
participants' guesses, computes the group result, presents that result, and
begins the next round with the updated context.

Execution produces four linked data layers. An \texttt{Ask} call returns
parsed responses keyed by participant ID. Procedure combines these responses
with conditions, item or round identifiers, and computed outcomes to produce
study data conforming to the Response specification.
\texttt{measurement.py} analyzes these data to compute study-specific measures
and statistical comparisons. In parallel, the runtime records an execution
trace containing the participant configuration and, for each call, its
assigned condition and role, visible context, raw model output, parsed
response, and resulting study state. Identifiers link the analyzed study
records to the calls that produced them.

\section{Authoring Tool Implementation}
\label{sec:system-reorganized}

\begin{figure*}[t]
    \centering
    \includegraphics[width=\textwidth]{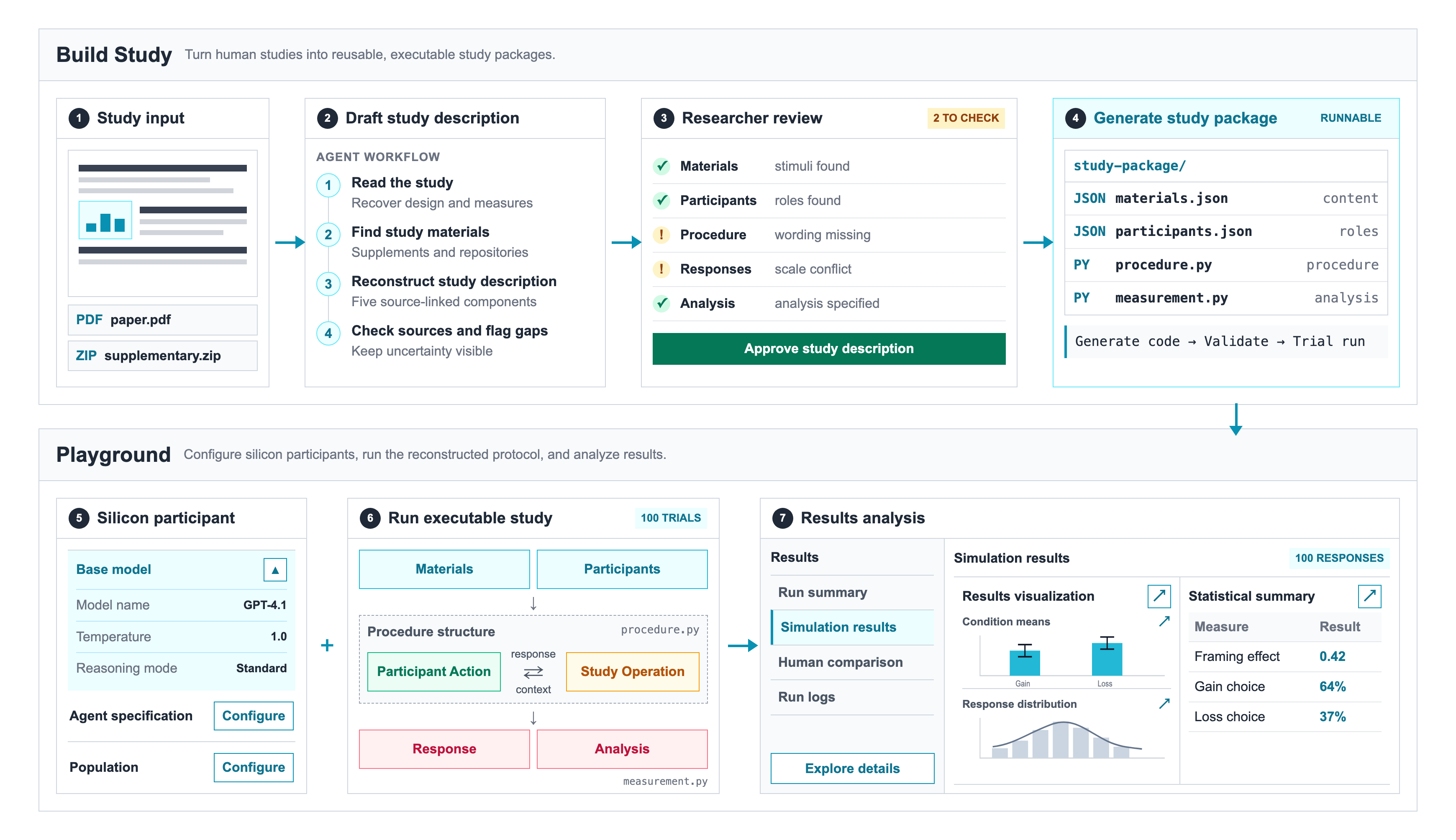}
    \caption{\textbf{Detailed Tool Workflow.} In \emph{Build Study}, researchers reconstruct and
    review a source-linked study description, which the tool turns into an
    executable \textsc{HumanStudyProgram}. In \emph{Playground}, researchers configure
    silicon participants, run the program, and examine its results from
    study-level findings to individual responses.}
    
    \label{fig:system-workflow-reorganized}
\end{figure*}

Sections~\ref{sec:study-abstractions} and~\ref{sec:runtime} describe the study
abstractions and their runtime. The tool exposes these capabilities through
two linked stages (Figure~\ref{fig:system-workflow-reorganized}): \emph{Build
Study} supports study reconstruction and review, while \emph{Playground}
supports participant configuration, execution, and result analysis.

\subsection{\emph{Build Study}: Reconstructing and Reviewing a Study}
\label{subsec:reconstruct-review}

In \emph{Build Study}, a user supplies the available source materials,
including a paper, supplementary files, and repository links. An LLM-assisted
pipeline identifies the empirical studies described across these sources,
reconstructs their Materials, Participants, Procedure, Response, and Analysis,
and links each field to its source location.

The interface organizes fields requiring attention into a navigable review
list and presents each field alongside its source evidence. It highlights
missing, unclear, conflicting, or unsupported information rather than silently
completing it. Users can compare a field with its source, edit its value, add
missing information, resolve conflicting descriptions, or ask the pipeline to
revise a selected part. Saved revisions and review notes remain attached to
the study description, and unresolved issues remain visible until the user
decides whether it is ready for approval.

\begin{figure*}[t]
\centering
\includegraphics[width=\textwidth]{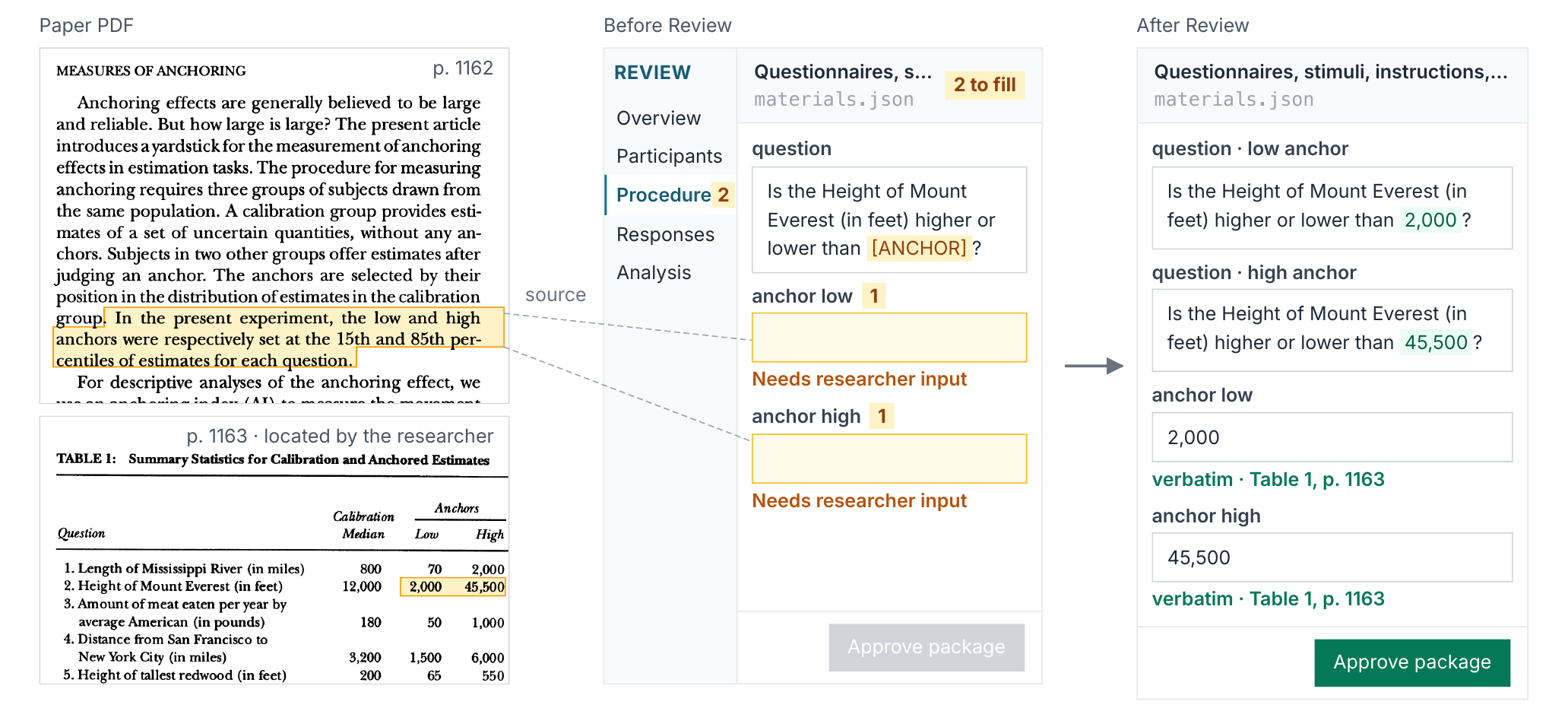}
\caption{\textbf{Recovering Missing Study Information.} The pipeline
recovers the passage defining how the anchors were set and flags the two
values it could not read from the scanned table. The researcher locates
Table~1 of the source paper and enters them; the package records each value as verbatim from
that table.}
\label{fig:case-anchoring-reorganized}
\end{figure*}

The anchoring study of Jacowitz and Kahneman~\citep{Jacowitz1995MeasuresOA}
illustrates this review process. In this experiment, participants estimate
unknown quantities after being exposed to either a high or low numerical
anchor. The reconstruction pipeline recovered the study structure but failed
to recover several anchor values from the older PDF. Rather than supplying
plausible values, the interface marked the missing information and placed it
alongside the source passages that specified the percentile rules
(Figure~\ref{fig:case-anchoring-reorganized}). The user could enter the
verified values and approve the study description while retaining their source references.

\subsection{\emph{Playground}: Configuring and Executing Silicon Participants}
\label{subsec:configure-execute}

\begin{figure*}[t]
    \centering
    \includegraphics[width=\textwidth]{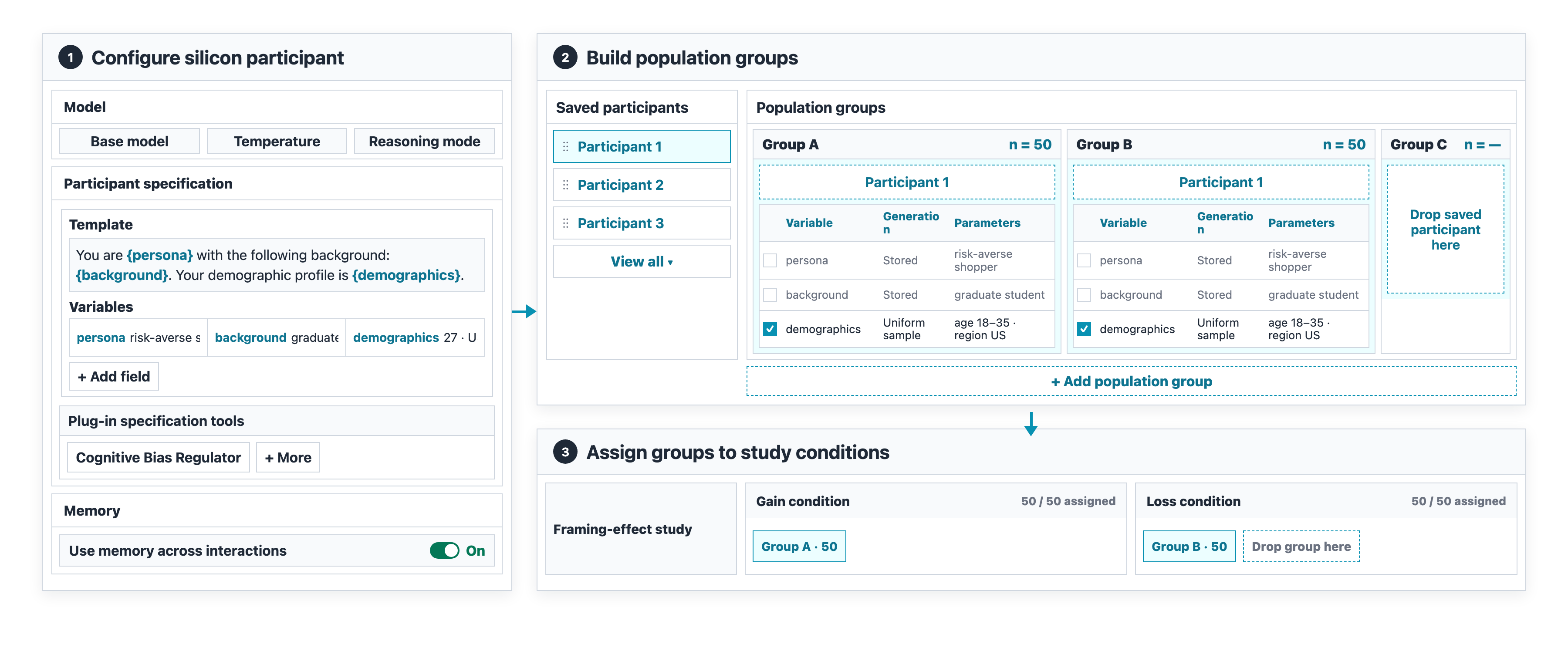}
    \caption{\textbf{Silicon Participant Configuration.} Users
    configure reusable silicon participants, form population groups, and
    assign them to study conditions.}
    
    \label{fig:silicon-participant-config-reorganized}
\end{figure*}

The \emph{Playground} stage runs an approved study package with silicon
participants. Users define reusable participants through a model, participant
specification, and memory, following the participant abstraction in
Figure~\ref{fig:intro-participant-abstraction}(a). Participant specifications
can also incorporate optional plug-in tools, such as a cognitive bias
regulator~\citep{Cobra}, to configure agent behavior. Users then form
population groups from saved participants, choose which specification
variables remain fixed or vary within a group, and set the group size
(Figure~\ref{fig:silicon-participant-config-reorganized}).

For execution, users select an empirical substudy and assign configured groups
to its conditions or roles. They can use explicit assignments or a
random-allocation policy and vary exposed trial-level parameters for
exploratory runs. \emph{Playground} records these choices as run configuration
rather than modifying the approved study program, making clear which elements
were held fixed and which were varied across runs.

\begin{figure*}[t]
\centering
\includegraphics[width=\textwidth]{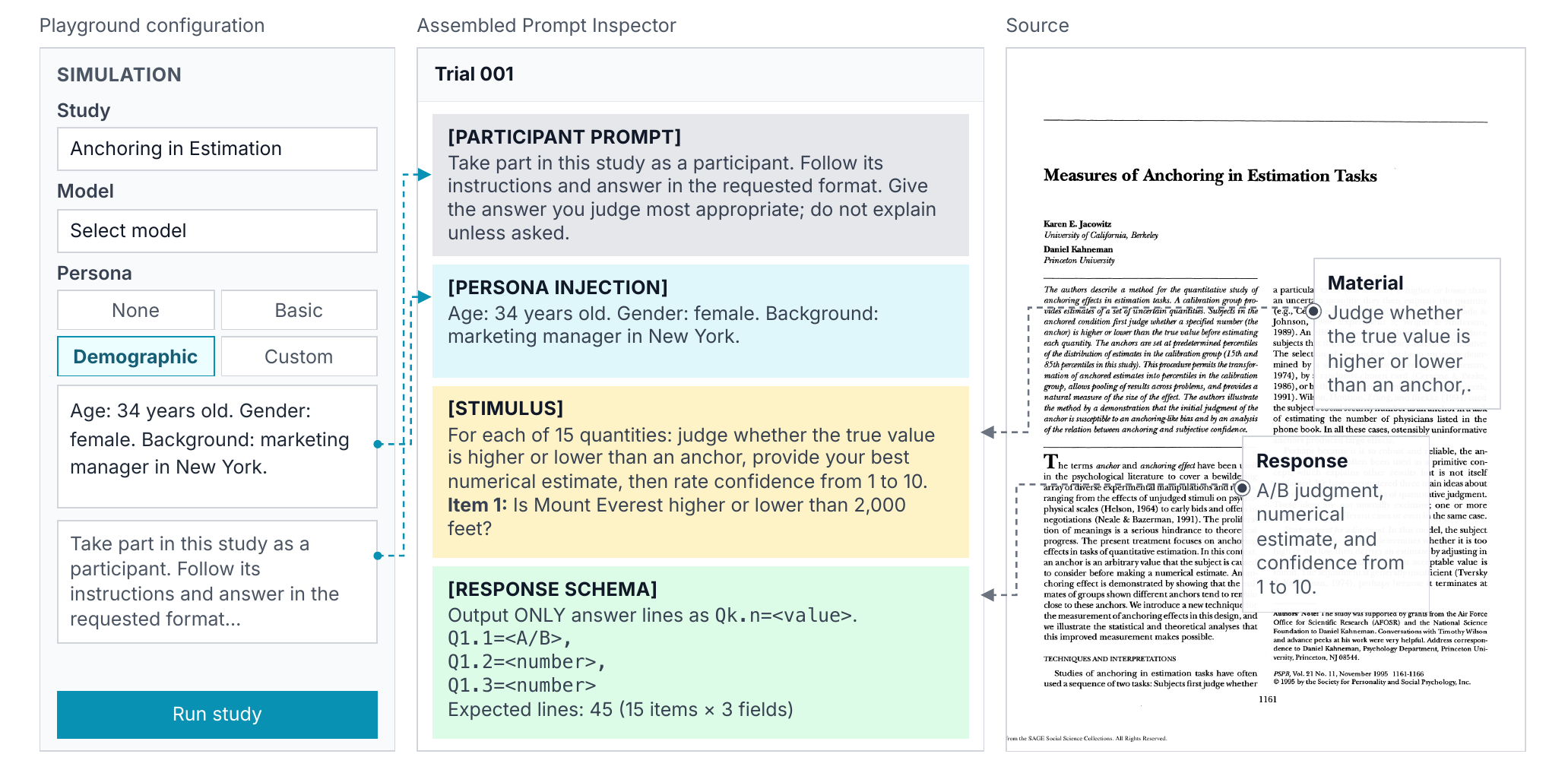}
\caption{\textbf{Tracing Silicon Participant Context.} The assembled context
distinguishes the participant configuration from the study content and links
the latter to its source.}
\label{fig:case-context-reorganized}
\end{figure*}

The anchoring study also illustrates how the system makes the context of a
silicon-participant response visible. A user can add a demographic
persona to the run configuration while leaving the approved study content
unchanged. Before execution, the interface separates the participant template
and persona from the stimulus and response schema supplied by the package
(Figure~\ref{fig:case-context-reorganized}). The source view links the study
content to its source passage. The same separation is retained with each
response, making clear which input belongs to the reconstructed study and
which belongs to the participant used for that run.

\subsection{\emph{Playground}: Exploring Results}
\label{subsec:analyze-records}

\begin{figure*}[t]
\centering
\includegraphics[width=\textwidth]{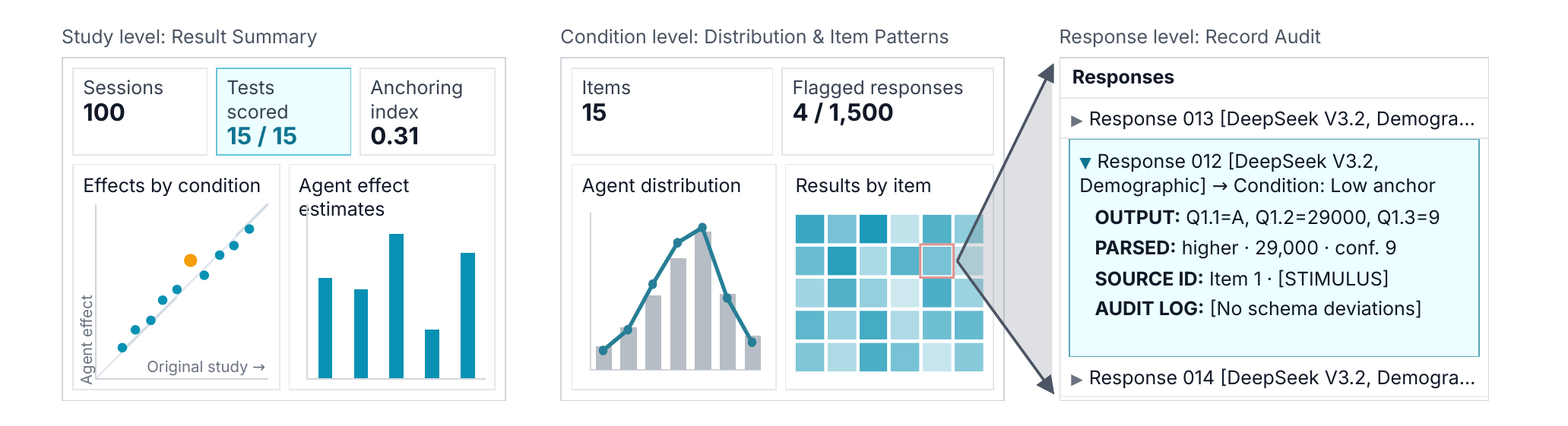}
\caption{\textbf{Multilevel Result Analysis.} Three linked views present the
results of a run: the study level summarizes findings and effects by condition,
the condition level shows response distributions and item-level patterns, and
the response level opens an individual record with its raw output, parsed
values, and audit log.}
\label{fig:analysis-levels-reorganized}
\end{figure*}

\emph{Playground} presents the study data, measurement outputs, and execution
traces produced by the runtime through linked views
(Figure~\ref{fig:analysis-levels-reorganized}). Users can move from a
study-level finding to its condition-level values, response distributions, and
statistical tests, then open a contributing response to inspect its participant
configuration, visible context, raw model output, and parsed response.

\begin{figure*}[t]
\centering
\includegraphics[width=\textwidth]{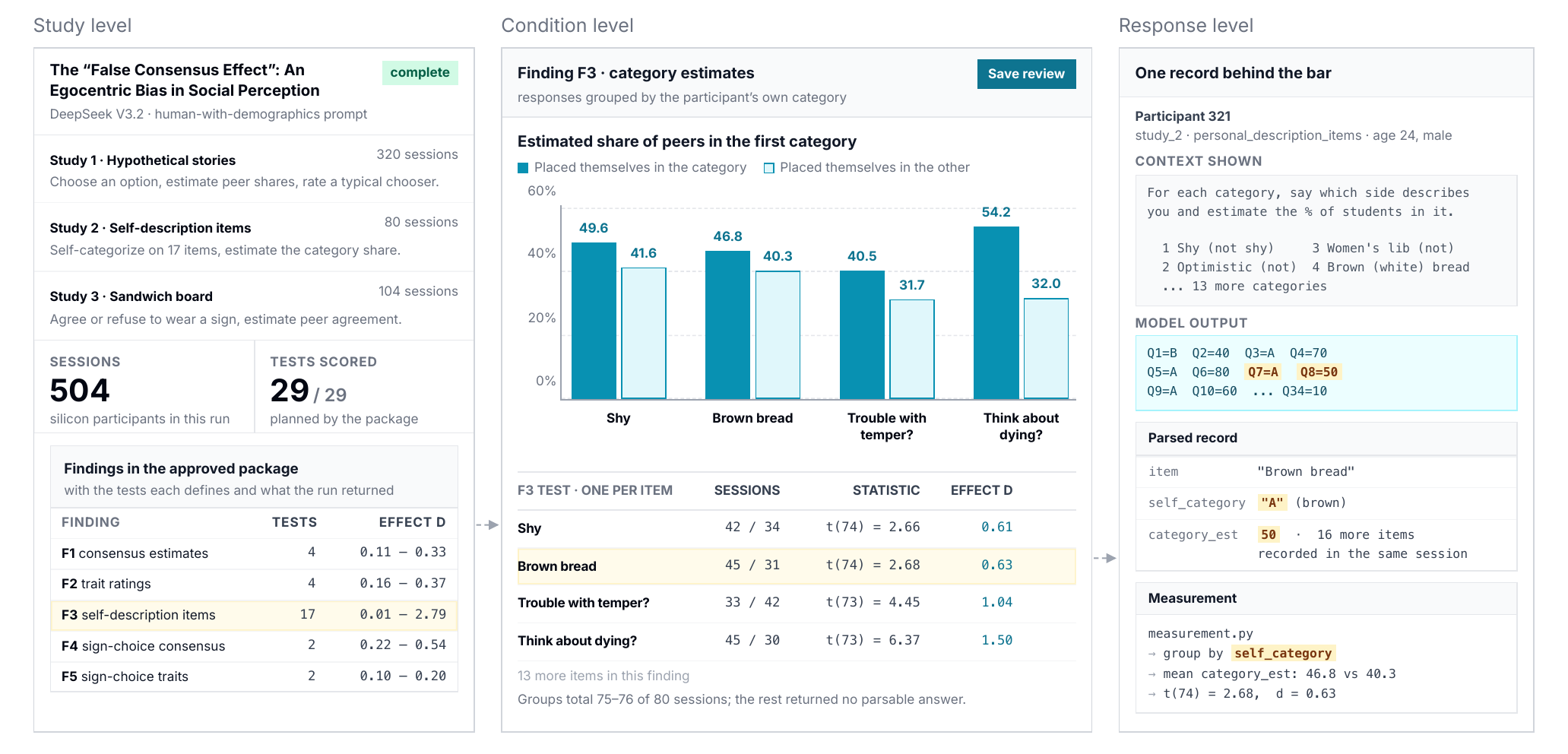}
\caption{\textbf{Following a Finding to Its Records.} A study-level finding
leads to its condition-level values and statistical tests, and then to an
individual response record and the measurement steps that consume it.}
\label{fig:case-results-reorganized}
\end{figure*}

The false-consensus study of Ross, Greene, and
House~\citep{Ross1977TheC} provides a concrete walkthrough. Participants place
themselves on one side of a choice or personal description and estimate the
share of their peers in a given category. The analysis compares that estimate
between the participants who placed themselves in the category and those who
did not. Figure~\ref{fig:case-results-reorganized}
follows one resulting finding across three views. The study-level view lists
the findings and tests computed for the run. Selecting a finding opens the
condition-level quantities and statistics behind it. Selecting one of these
groups opens a contributing response record, including the participant's
visible context, raw model output, and parsed response. The accompanying
measurement view identifies the fields consumed by
\texttt{measurement.py} and how they are reduced to the displayed statistic.
The views aggregate the same study data at different levels and link back
to the corresponding execution trace.

\section{Expert Review}
\label{sec:usability}

We conducted a hands-on expert review to understand how \textsc{HumanStudyProgram} could support social science research practice. Each participant used a familiar study from their area of expertise, allowing them to assess its reconstruction and execution against the source materials and their domain knowledge. We observed how they reviewed and corrected the reconstructed study, configured and ran silicon participants, and interpreted the results. We also asked what value and limitations they saw and how they would use the tool in their own research.

\subsection{Participants}

We purposively recruited 12 doctoral-level social science researchers. Participants were pursuing or had completed doctoral training, had more than three years of experience conducting human-subjects research, and represented economics, psychology, political science, and management. All participants had experience with both quantitative and qualitative methods, as well as prior experience with LLMs and LLM-based silicon participants. Each participant received \$40 for an approximately 70-minute session.

\subsection{Study Design and Tasks}

Each participant selected a human-subject study from their area of expertise that had not previously been processed by the tool. Some selected their own papers or papers by close collaborators; others selected studies they had examined closely. Their familiarity gave them a concrete basis for evaluating how well the tool represented and executed the study.

Participants used the complete workflow. In Build Study, they compared the reconstructed study with the submitted source materials and their own knowledge. They corrected inaccurate or missing information, addressed issues identified by the tool, and approved the reconstructed study description. The system then generated
and validated its executable package. In Playground, they configured silicon participants by selecting a model, defining participant profiles, setting the sample size, and adjusting other run settings. They then ran the study and examined the results at the study, condition, and individual-response levels. Participants could freely edit the reconstructed study, choose different simulation settings, and conduct multiple runs.

Participants submitted their papers and any public supplementary materials before the session. We completed and cached the extraction in advance so that the session focused on hands-on use and evaluation of the tool. We used Gemini 3 Flash for these extractions after formative testing of four models; Appendix~\ref{app:build-study-model-selection} describes this selection. Appendix~\ref{app:study-sample} reports extraction and review times for the submitted studies, and Appendix~\ref{app:expert-run-configurations} reports the models, number of runs, and configured silicon-participant populations used during the sessions.

\subsection{Procedure}

Sessions were conducted online through a videoconferencing platform. Each session involved one participant, one interviewer, and one note-taker. Participants kept their cameras on and shared their screens while using the deployed web interface on their own computers. Each session consisted of three stages.\\
\textbf{Tutorial (approximately 10 minutes).} The session began with a brief background discussion and a tutorial using the same cached example. The interviewer introduced the main workflows of \textsc{HumanStudyProgram} and answered questions about interface operation.\\
\textbf{Tool Use and Think-Aloud (approximately 40 minutes).} Participants used the tool with their selected study. We encouraged them to think aloud by describing what they noticed, expected, and considered while making decisions. Based on these comments, the interviewer asked neutral follow-up questions to clarify the reasons behind participants' actions and judgments. The interviewer also answered questions about interface operation without influencing how participants used the tool or evaluated its output.\\
\textbf{Interview and Survey (approximately 20 minutes).} The session concluded with a semi-structured interview about the participant's experience, possible uses of the tool, and its limitations. Participants then completed the post-study survey. The think-aloud instructions, interview guide, and survey are provided in Appendix~\ref{app:researcher-study}.

\subsection{Data Collection and Analysis}

With consent, we recorded participants' screens and audio and transcribed their think-aloud comments and interviews. A note-taker documented participants' actions, questions, and comments. We also collected post-study survey responses.

We summarized survey responses on their original seven-point scales. Two researchers independently analyzed the complete set of think-aloud and interview transcripts from all 12 participants. Each researcher marked relevant passages and identified recurring patterns across participants. They then compared their analyses, discussed differences in interpretation, and jointly developed and agreed on the final themes concerning research value, workflow use, areas for improvement, and boundaries of appropriate use. We used the screen recordings and notes to relate participants' comments to their actions and considered cases that differed from the main patterns.

\subsection{Ethical Considerations}

The study received institutional ethics approval. Participants provided informed consent and could stop at any time. We collected only demographic information relevant to the analysis. Recordings and transcripts were stored in access-controlled research storage. We refer to participants only by identifiers in the paper.

\subsection{Results}

We refer to participants as P1 through P12. In Build Study, all 12 participants reviewed the reconstruction of their selected study, addressed the issues requiring their judgment, and approved a study description from which the system generated an executable package. Across the 12 reconstructions, the tool surfaced 98 issues for researcher confirmation, with between 4 and 15 issues per study (Appendix~\ref{app:study-sample}). Reconstruction itself ran without the researcher and took 10 to 30 minutes per study; reviewing and approving the resulting package took a participant approximately 20 minutes, ranging from 10--15 to 25--30 minutes across the 12 studies. In Playground, all 12 participants configured silicon participants, executed the generated study package, and examined results at the study, condition, and individual-response levels. We organize the findings around potential uses, workflow and usability, and post-study ratings.

\subsubsection{Potential Uses}

Participants described three broad ways they would use the tool: understanding published study protocols, prototyping and piloting studies, and applying human study protocols to understand AI-agent behavior. Table~\ref{tab:envisioned-applications} summarizes the applications participants described.\\
\textbf{Understanding Published Study Protocols.} Participants saw the tool as a way to understand how a published study was actually conducted. The structured study components helped them connect the paper's description with its stimuli, conditions, measures, and analyses, making the protocol easier to inspect, reproduce, and use for teaching (P3, P6, P8, P11). As P8 explained, \textit{``I would really like to use this kind of structured extraction to help me read papers.''}\\
\textbf{Prototyping and Piloting Studies.} All participants wanted to use the tool before collecting human data. They described modifying conditions in existing protocols, testing alternative designs, adapting studies to new questions or contexts, and importing their own study designs to find problems before running a human study. As P3 explained, \textit{``I would use it after I have an idea for an experiment, but before I run the human study. It could help me see what is missing and whether the design works as I expected.''}\\
\textbf{Applying Human Study Protocols to Understand AI-Agent Behavior.} Participants also wanted to apply established human study protocols to AI agents. By holding the protocol fixed and varying models, prompts, personas, or languages, they could examine agent preferences and biases and compare AI and human behavior (P2, P4, P7, P8, P11, P12). As P2 explained, \textit{``AI agents themselves have now become subjects of social science research.''}

\begin{table*}[t]
    \centering
    \footnotesize
    \setlength{\tabcolsep}{8pt}
    \renewcommand{\arraystretch}{1.15}
    \caption{Applications identified by participants across the three potential uses of the tool.}
    \label{tab:envisioned-applications}
    \begin{tabularx}{\textwidth}{@{}X|X|X@{}}
        \toprule
        \textbf{Understanding Protocols}
        & \textbf{Prototyping and Piloting}
        & \textbf{Studying AI Agents} \\
        \midrule
        Reconstructing and rerunning study protocols
        & Piloting studies before data collection
        & Understanding AI-agent behavior \\
        Helping researchers understand papers
        & Finding problems before data collection
        & Comparing AI and human behavior \\
        Practicing study-design skills
        & Testing alternative study designs
        & \\
        Teaching with executable studies
        & Investigating new research questions
        & \\
        & Adapting studies to new contexts
        & \\
        \bottomrule
    \end{tabularx}
\end{table*}

\subsubsection{Workflow and Usability Findings}
We first report participants' overall experience with the workflow, followed by their experiences with study reconstruction, human review and correction, configuring and executing silicon participants, and interpreting results.\\
\textbf{Overall Workflow Experience.} Many participants valued having study reconstruction, agent configuration, execution, and result analysis connected in one workflow. This reduced the need to coordinate separate tools and made repeated runs across models and configurations more practical. As P11 explained, \textit{``This is more or less automated, the whole workflow. You can do a lot of iterations with different combinations, different iterations, different models.''} Similar views were expressed by P1, P4, P5, P8, and P10.\\
\textbf{Study Reconstruction.} Participants valued how the tool's structured extraction organized a paper into familiar research components, such as the study overview, stimuli, measures, and analyses. This structure helped them quickly locate the information they cared about and understand how a study could be reproduced (P11). Participants generally perceived the extraction as accurate, particularly in its representation of experimental variables, procedures, and analyses. As P3 noted, \textit{``The independent variables, dependent variables, and how the data were analyzed were all extracted very clearly.''} Several participants felt that the tool's structure better reflected what researchers look for when reading a paper than general-purpose AI tools. As P7 explained, \textit{``Your toolkit understands what I want better than my AI chatbot. With ChatGPT, I have to explain what I am looking for. Sometimes, even after many rounds of conversation, it still cannot give me the part of the paper I need, and I end up reading the paper again myself.''} Others made similar comparisons with Codex and Claude Code (P4, P6). Participants also valued being able to navigate directly to specific study components rather than search through the entire paper (P2, P3, P8, P11).\\
\textbf{Human Review and Correction.} The tool surfaced missing, inconsistent, and uncertain details and showed which decisions required researcher judgment. P11 described the explanation provided for each issue as \textit{``very comprehensive''} and helpful for deciding how to address it. In one case, the tool flagged a mismatch between the scale described in a paper's method section and the scale used in the experiment, then suggested contacting the authors. As P7 explained, \textit{``Your toolkit found an error in the paper and suggested contacting the authors. The scale described in the method section did not match the one used in the experiment. When I read the paper carefully, I also felt that something was not quite right.''} 

Participants also valued that missing-information alerts led directly to the relevant source text and the item requiring confirmation (P3, P6, P7, P11). However, P5 and P9 found the amount of information demanding to review; each spent 20--25 minutes in \emph{Build Study} and suggested a mind-map view organizing the reconstructed study components and
their relationships to help them navigate the information during review.\\
\textbf{Configuring and Executing Silicon Participants.} Participants tested the tool with familiar papers spanning different languages, stimuli, participant configurations, procedures, and analyses. P2 was surprised that a bilingual study gave each participant the appropriate stimuli based on their language. P4 similarly valued how the tool handled multiple participants in the same scenario and recovered the study's multi-turn structure without requiring them to reconstruct the sequence of prompts. 

Participants valued being able to see what silicon participants actually received. Before starting a run, they could review the system and user prompts, how stimuli were presented, and how multi-turn procedures were translated into agent interactions. This visibility reduced participants' concerns about relying on AI agents and made them more comfortable using the tool to run experiments. As P1 explained, \textit{``I can see exactly what the agent sees. That greatly reduces my distrust of AI agents and makes me feel more comfortable using your toolkit to run an experiment.''} 

The tool also showed the model, participant profiles, sample size, language, and reasoning settings for each run. Participants said these records helped them understand how a run was produced rather than treating it as a black box. P8 viewed them as research material for comparing runs and understanding why agents behaved differently, while P11 valued using them for systematic comparisons across models and configurations. However, P2 and P3 found the configuration process difficult to navigate and wanted clearer guidance and a simpler interface.\\
\textbf{Interpreting Results.} Participants moved from the overview to condition-level results and response distributions, then to individual prompt--response records when a result required further explanation. This connected summary findings to the evidence behind them. As P9 explained, \textit{``I did not expect that I could click into every level and keep going deeper. Being able to move from the overall result to the details behind it was very helpful for understanding where the result came from.''} Similar navigation was observed for P4, P8, and P11. Participants used visualizations and statistical results to identify patterns, and individual responses and recorded prompts to examine why those patterns emerged (P4, P9, P11).

\subsubsection{Post-Study Ratings}

All seven item medians were above the scale midpoint and reflected the qualitative findings. The tool's value for research practice received the strongest ratings (median 6, IQR 6--7), with every participant selecting 6 or 7. Ratings for study match and clarity of the simulation context also had medians of 6; clarity ratings were the most tightly clustered, ranging from 5 to 6. Ease ratings had medians of 5. Reviewing and revising the reconstructed study and study match shared the widest range (4--7), while configuring a run ranged from 4 to 6, consistent with participants' requests for a clearer overview and configuration guidance. Table~\ref{tab:user-study-ratings} reports the complete distribution summaries, and the complete items are provided in Appendix~\ref{app:researcher-study}.

\begin{table}[t]
    \centering
    \small
    \caption{Post-study survey responses ($N=12$) on the original seven-point scales. Higher values indicate greater ease, match, clarity, or value; 4 is the scale midpoint. IQR reports the first-to-third-quartile interval.}
    \label{tab:user-study-ratings}
    \begin{tabularx}{0.92\columnwidth}{@{}Xccc@{}}
        \toprule
        Item & Median & IQR & Range \\
        \midrule
        Ease of using the complete workflow & 5 & 5--6 & 5--7 \\
        Match with the selected study & 6 & 6--6 & 4--7 \\
        Ease of reviewing and revising the reconstructed study & 5 & 5--6 & 4--7 \\
        Clarity of the simulation context & 6 & 6--6 & 5--6 \\
        Ease of interpreting simulation results & 5 & 5--6 & 5--7 \\
        Ease of configuring a simulation run & 5 & 5--5 & 4--6 \\
        Value for research practice & 6 & 6--7 & 6--7 \\
        \bottomrule
    \end{tabularx}
\end{table}

\section{Research Deployment}
\label{sec:deployment}

We deployed \textsc{HumanStudyProgram} as a web application and shared it with the social science research community. Within one week, 23 research teams contacted us about using the tool and potential collaboration. We established ongoing collaborations with four teams pursuing different research goals. Together, they have converted over a hundred papers into executable study protocols. A psychology team converted studies of human ethical judgment into executable protocols to examine how AI agents respond to ethical questions. A management research team converted human studies in international business into executable protocols to examine whether AI agents could serve as valid silicon participants in the field. A team studying tsunami evacuation connected empirical evidence about human behavior with agent-based simulations to model individual and collective decisions. An operations research team used the tool to prototype and debug decision-making experiments before data collection. At the time of writing, all four collaborations had continued for more than two months, and each team had begun a new research project supported by \textsc{HumanStudyProgram}.

\section{Discussion}
\label{sec:discussion}

\textbf{Human Study Protocols as Instruments for Studying AI Agents.} Human study protocols bring established behavioral theories into the study of AI agents~\citep{rahwan2019machine,binz2023using}. Whether theories developed to explain human behavior also explain AI behavior remains an open question~\citep{yax2024reasoning}. Agreement with a human finding can support further investigation, while differences can reveal where a theory may need revision. Both are informative when agents themselves are the subject of research. \textsc{HumanStudyProgram} could support this work across multiple tasks and settings, providing a basis for developing and testing theories of AI behavior.\\
\textbf{Aligning Agent Specifications with Observed Behavior.} Behavioral evidence could also help researchers refine agent specifications. A specification describes intended characteristics, but its value depends on whether those characteristics are reflected in behavior. \textsc{HumanStudyProgram} could connect these descriptions to observations across studies, allowing researchers to revise a specification and evaluate the resulting changes. Testing on additional studies would help determine whether an adjustment generalizes beyond the tasks used to make it~\citep{yarkoni2022generalizability,binz2025foundation}. Agent specifications could then be supported by evidence of how agents behave and the settings in which that behavior has been evaluated.\\
\textbf{Understanding Community Norms in Research Writing.} Reconstructing studies could also help us understand how research communities write about their work. Papers communicate more than procedures: they frame questions, justify design choices, and connect evidence to conclusions. Comparing how papers present a study with how its components fit together could reveal conventions in this process, including what authors emphasize and what they expect readers to infer. Extending \textsc{HumanStudyProgram} to examine these relationships across papers could help characterize how such conventions differ across fields. This could support researchers learning to write for a new community and tools that help authors explain the reasoning behind a study. It could also make those conventions easier to question, including whether familiar ways of presenting research leave important decisions or alternative interpretations unexplained.\\
\textbf{Toward Executable Research Publications.} The same representation could also change how research papers are produced. Today, \textsc{HumanStudyProgram} reconstructs a study from a paper. A future workflow could reverse this process and generate a paper from the study design, execution records, data, and analyses. The system could draft the manuscript using these records and the authors' account of the research question, while authors check whether its explanations and conclusions are supported by the study. Readers could access both the paper and the executable study, follow results back to the underlying analyses, and run modified versions to explore new questions. The publication could then become a written account of a study that remains available for further work. Automatic paper generation would be part of this broader change, keeping the manuscript and the study it describes connected as both are revised.\\
\textbf{Scope and Limitations.}
\textsc{HumanStudyProgram} currently focuses on text-based social experiments,
including single- and multistage procedures, repeated interactions, and
role-based multi-participant studies. It does not yet support studies that
depend on physical action, physiological sensing, or immersive and other
rich-media environments. The quality of a generated study program depends on
the information available in the paper and its supplementary materials.
Silicon-participant behavior reflects the capabilities and characteristics of
the underlying LLM. The resulting runs support protocol inspection,
early-stage prototyping, and research on AI behavior, while claims about human
populations continue to rely on human-participant evidence.
\section{Conclusion}
\label{sec:conclusion}

We presented \textsc{HumanStudyProgram}, a researcher-facing framework that
reconstructs a documented human study as an executable program while separating the study
procedure from the silicon participants used to run it. We implemented a
researcher-facing authoring tool that supports source-linked study reconstruction and
review, configurable silicon-participant execution, and navigation from
aggregate findings to individual responses and their execution context.
Through study demonstrations and a hands-on review with 12 social science
researchers, we found value in understanding published protocols, prototyping
studies, and examining AI behavior. By making documented study procedures
executable and reusable, we hope \textsc{HumanStudyProgram} can make human
study methods easier to reuse and build upon.

{\small\renewcommand{\refname}{\normalsize\sffamily\bfseries References}
\bibliographystyle{unsrtnat}
\bibliography{references}}

\appendix
\section{Build Study Model Selection}
\label{app:build-study-model-selection}

During development, we evaluated the ability of Gemini 3 Flash, GPT 5 Nano, Claude Haiku 4.5, and DeepSeek V3.2 to extract and build structured study packages on 10 older papers published between 1946 and 2007 and 20 papers published after 2015. The papers covered psychology, economics, political science, and management. Older papers often contained scanned pages, tables, and text that was difficult to recover with OCR. Newer papers more often distributed study information across longer articles, supplementary files, and external repositories. All four models produced schema-conforming candidate packages and surfaced similar types of omissions and uncertainties. Their main difference was how they described the extracted content and confirmation issues.

We used Gemini 3 Flash in the expert review because its descriptions were clearer and more natural for researchers to review, while its processing cost remained practical. Table~\ref{tab:build-study-model-selection} summarizes this formative comparison. Since the four models produced similar reconstruction outcomes, we compare their use in the pipeline rather than rank their accuracy.

\begin{table}[t]
    \centering
    \small
    \setlength{\tabcolsep}{4pt}
    \renewcommand{\arraystretch}{1.12}
    \caption{Model selection criteria and observations for the \emph{Build Study} pipeline.}
    \label{tab:build-study-model-selection}
    \begin{tabularx}{0.84\columnwidth}{@{}>{\raggedright\arraybackslash}p{0.25\columnwidth}>{\raggedright\arraybackslash}X@{}}
        \toprule
        \textbf{Criterion} & \textbf{Observation} \\
        \midrule
        Pipeline completion & All four models generated structured study packages for researcher review. \\
        \addlinespace[2pt]
        Omissions and uncertainties & The four models surfaced broadly similar types of issues. \\
        \addlinespace[2pt]
        Presentation & The models differed mainly in how they described reconstructed content and confirmation issues. Gemini 3 Flash produced the clearest and most natural descriptions. \\
        \addlinespace[2pt]
        Operational cost & Gemini 3 Flash provided a practical processing cost for complete study materials. \\
        \addlinespace[2pt]
        Final selection & We used Gemini 3 Flash for \emph{Build Study} in the expert review. \\
        \bottomrule
    \end{tabularx}
\end{table}

\section{Researcher Study Materials}
\label{app:researcher-study}

\subsection{Selected Studies and Submitted Materials}
\label{app:study-sample}

The selected studies covered economics (6), psychology (2), political science (2), and management (2). Five were submitted as PDFs and seven as ZIP archives containing supplementary materials. A confirmation issue is an independently recorded item that required the researcher to verify information, supply missing information, resolve ambiguity or conflict, or accept a documented departure from the source study. Extraction time is the duration of the automated reconstruction, which we completed and cached before the session and which required no researcher involvement. Review time is the part of the hands-on block a participant spent in \emph{Build Study}, from opening the reconstructed study to approving the study description. We estimated it from the screen recordings and report it in five-minute bands; it excludes the tutorial, the \emph{Playground} configuration and runs, and the closing interview. Estimated JSON tokens characterize the amount of structured study information in the metadata, materials, and task specification. We divide the character count of normalized JSON by four and round to the nearest hundred tokens.

\begin{table*}[t]
    \centering
    \small
    \setlength{\tabcolsep}{3.5pt}
    \renewcommand{\arraystretch}{1.08}
    \caption{Participant-selected studies, submitted materials, approximate extraction and review times, and issues surfaced for researcher confirmation ($N=12$).}
    \label{tab:selected-study-summary}
    \begin{tabularx}{\textwidth}{@{}lc>{\raggedright\arraybackslash}Xlcccc@{}}
        \toprule
        Study & Year & Domain & Materials & \shortstack{Extraction\\time} & \shortstack{Review\\time} & \shortstack{Estimated\\JSON tokens} & \shortstack{Confirmation\\issues} \\
        \midrule
        P1  & 2016 & Management & PDF & 10--15 min & 20--25 min & 6,200 & 10 \\
        P2  & 2013 & Psychology & PDF & 10--15 min & 20--25 min & 8,100 & 11 \\
        P3  & 2024 & Psychology & ZIP & 25--30 min & 15--20 min & 4,000 & 7 \\
        P4  & 2016 & Economics & ZIP & 25--30 min & 15--20 min & 5,300 & 5 \\
        P5  & 2024 & Economics & ZIP & 25--30 min & 20--25 min & 6,200 & 11 \\
        P6  & 2024 & Political science & PDF & 10--15 min & 25--30 min & 8,500 & 15 \\
        P7  & 2026 & Political science & ZIP & 25--30 min & 15--20 min & 5,200 & 7 \\
        P8  & 2021 & Management & ZIP & 25--30 min & 15--20 min & 1,500 & 7 \\
        P9  & 2023 & Economics & ZIP & 25--30 min & 20--25 min & 9,200 & 6 \\
        P10 & 2011 & Economics & ZIP & 25--30 min & 10--15 min & 1,500 & 4 \\
        P11 & 1995 & Economics & PDF & 10--15 min & 15--20 min & 6,700 & 6 \\
        P12 & 2002 & Economics & PDF & 10--15 min & 20--25 min & 6,400 & 9 \\
        \bottomrule
    \end{tabularx}
\end{table*}

\subsection{Expert Review Run Configurations}
\label{app:expert-run-configurations}

Table~\ref{tab:expert-run-configurations} summarizes the simulation runs completed during the hands-on sessions. DeepSeek V3.2 was the default model presented to experts when configuring silicon participants. They could retain it or select another available model, including Gemini 3 Flash, Claude Haiku 4.5, and GPT 5 Nano, and could conduct additional runs as time allowed. Population reports the total number of silicon participants configured for each run.

\begin{table}[t]
    \centering
    \small
    \caption{Models, runs, and configured silicon-participant populations in the expert review ($N=12$).}
    \label{tab:expert-run-configurations}
    \begin{tabularx}{\columnwidth}{@{}lXcc@{}}
        \toprule
        Participant & Model used & Runs & Population per run \\
        \midrule
        P1  & DeepSeek V3.2 & 1 & 20 \\
        P2  & DeepSeek V3.2 & 1 & 30 \\
        P3  & DeepSeek V3.2; Gemini 3 Flash & 2 & 50 \\
        P4  & DeepSeek V3.2; Claude Haiku 4.5 & 2 & 12 \\
        P5  & DeepSeek V3.2 & 1 & 50 \\
        P6  & DeepSeek V3.2; Gemini 3 Flash; GPT 5 Nano & 3 & 100 \\
        P7  & DeepSeek V3.2 & 1 & 40 \\
        P8  & DeepSeek V3.2; Gemini 3 Flash; Claude Haiku 4.5 & 3 & 50 \\
        P9  & DeepSeek V3.2; Gemini 3 Flash & 2 & 20 \\
        P10 & DeepSeek V3.2 & 1 & 30 \\
        P11 & DeepSeek V3.2 & 1 & 30 \\
        P12 & DeepSeek V3.2 & 1 & 40 \\
        \bottomrule
    \end{tabularx}
\end{table}

\subsection{Think-Aloud Instructions}

Before participants began using the tool with their selected study, the interviewer provided the following instructions:

\begin{quote}
For the next part of the session, please use the tool with your selected study. You may work through the tool in whatever way makes sense to you. As you work, please say aloud what you are looking at, what you are trying to do, what you expect to happen, and what you are considering when making a decision. Please describe your reactions as they occur.

There are no right or wrong actions or answers. We are studying the tool and are not evaluating your performance. You may decide what to review, change, configure, or run. We will not direct those choices. You may ask questions about how the interface works at any time. We can explain its operation. Decisions about how to use the tool remain yours.

If you become quiet, I may remind you to keep talking or ask what you are thinking. Do you have any questions about this process before we begin?
\end{quote}

During the session, the interviewer used only neutral prompts when clarification was needed, including ``Please keep talking,'' ``What are you thinking now?,'' ``What are you trying to do?,'' ``What are you noticing?,'' ``What did you expect to happen?,'' and ``What led you to that choice?''

\subsection{Post-Study Survey}
\label{app:post-study-survey}

Participants answered seven questions. Each used a seven-point response scale with labeled endpoints. Responses were coded from 1, representing the first endpoint listed below, to 7, representing the second endpoint.

\begin{enumerate}
    \item How easy or difficult was it to use the complete workflow? \emph{Very difficult} to \emph{very easy}.
    \item To what extent did the reconstructed study match your understanding of the selected study? \emph{Not at all} to \emph{completely}.
    \item How easy or difficult was the process of reviewing and revising the reconstructed study? \emph{Very difficult} to \emph{very easy}.
    \item How clear or unclear was the information about what silicon participants received? \emph{Very unclear} to \emph{very clear}.
    \item How easy or difficult was it to interpret the simulation results? \emph{Very difficult} to \emph{very easy}.
    \item How easy or difficult was it to configure a simulation run? \emph{Very difficult} to \emph{very easy}.
    \item Based on your experience, what value, if any, does the tool have for your research practice? \emph{No value at all} to \emph{substantial value}.
\end{enumerate}

\begin{table}[t]
    \centering
    \small
    \caption{Individual post-study survey responses. Q1--Q7 correspond to the seven survey questions listed in Appendix~\ref{app:post-study-survey}.}
    \label{tab:individual-survey-responses}
    \setlength{\tabcolsep}{5pt}
    \begin{tabular*}{\columnwidth}{@{\extracolsep{\fill}}lccccccc@{}}
        \toprule
        Participant & Q1 & Q2 & Q3 & Q4 & Q5 & Q6 & Q7 \\
        \midrule
        P1  & 6 & 7 & 5 & 6 & 5 & 5 & 7 \\
        P2  & 5 & 6 & 5 & 6 & 5 & 4 & 6 \\
        P3  & 5 & 6 & 6 & 6 & 5 & 4 & 7 \\
        P4  & 6 & 6 & 6 & 5 & 6 & 5 & 7 \\
        P5  & 5 & 6 & 4 & 6 & 5 & 5 & 6 \\
        P6  & 5 & 6 & 5 & 6 & 5 & 5 & 6 \\
        P7  & 5 & 6 & 6 & 6 & 5 & 5 & 6 \\
        P8  & 5 & 6 & 5 & 6 & 5 & 5 & 7 \\
        P9  & 5 & 6 & 4 & 6 & 6 & 5 & 6 \\
        P10 & 6 & 4 & 5 & 6 & 5 & 5 & 6 \\
        P11 & 7 & 6 & 7 & 5 & 7 & 5 & 7 \\
        P12 & 6 & 6 & 6 & 6 & 6 & 6 & 6 \\
        \bottomrule
    \end{tabular*}
\end{table}

\subsection{Semi-Structured Interview Guide}
\label{app:interview-guide}

The interviewer used the following questions. The order could follow the participant's actions. Neutral prompts such as ``Can you give an example?'' and ``What led you to that judgment?'' were used when clarification was needed.

\begin{enumerate}
    \item Please walk me through how you approached the task.
    \item What did you pay attention to during the workflow? Why?
    \item What, if anything, was easy, difficult, or unclear?
    \item How did you assess the reconstructed study?
    \item What did you change or leave unchanged? What informed those decisions?
    \item How did your knowledge of the selected study affect your review?
    \item How did you choose the run configurations? What did you do when viewing the results?
    \item How did you interpret the results? What evidence informed that interpretation?
    \item What uses, if any, could these artifacts and results support in your work?
    \item What conditions or concerns, if any, would affect whether you used these artifacts or results?
    \item What changes, if any, would improve the tool for your work?
    \item Is there anything else about your experience that you want to discuss?
\end{enumerate}

\end{document}